\documentclass[11pt,a4paper]{article}

\usepackage{amsfonts}
\usepackage{color}
\usepackage{graphicx}
\usepackage{float}

\usepackage{amsmath}
\usepackage{alltt}
\usepackage{cite}

\newcommand{\Eqref}[1]{(\ref{#1})}

\title{\textbf{Kink scattering in a generalized Wess-Zumino model}}

\author{A. Alonso-Izquierdo$^{(a,c)}$, M.A. Gonz\'alez Le\'on$^{(a,c)}$, J. Mart\'{\i}n Vaquero$^{(a,c)}$ \\
and M. de la Torre Mayado$^{(b,c)}$\\
$^{(a)}$ Departamento de Matematica Aplicada, Universidad de Salamanca \\
$^{(b)}$ Departamento de F\'{\i}sica Fundamental, Universidad de Salamanca \\
$^{(c)}$ IUFFyM, Universidad de Salamanca}

\date{}

\begin{document}

\maketitle

\begin{abstract}
In this paper, kink scattering in the dimensional reduction of the bosonic sector of a one-parameter family of generalized Wess-Zumino models with three vacuum points is discussed. The value of the model parameter determines the specific location of the vacua. The influence of the vacuum arrangements (evolving from three collinear vacua to three vacua placed at the vertices of an equilateral triangle) on the kink scattering is investigated. Two different regimes can be distinguished: in the first one, two symmetric BPS kinks/antikinks arise whereas in the second one a new different BPS kink/antikink emerges, with the exception of a three-fold rotational symmetry case, where the three topological defects are identical. The scattering between the two symmetric kinks is thoroughly analyzed. Two different scattering channels have been found: \textit{kink-kink reflection} and \textit{kink-kink hybridization}. In the last case, the collision between the two symmetric kinks gives rise to the third different kink. Resonance phenomena also appear allowing a vibrating kink to split into two symmetric kinks moving away.
\end{abstract}

\section{Introduction}

Wess-Zumino (WZ) models have drawn great attention in Physical Literature for the last decades. It has been considered as the first example of an interacting four-dimensional quantum field theory with linearly realized supersymmetry \cite{Wess1974}. A particularly interesting set of this family is given by the generalized WZ models with $\mathbb{Z}_N$ symmetry. The potential term in these models has $N$ vacua, which are equidistantly distributed in the unit circle centered at the origin of the complex internal space. In the supersymmetric context, the potential is derived from a superpotential of the form $W(\phi) = \phi- \frac{1}{N+1} \phi^{N+1}$ \cite{Cecotti1993, Alonso2000}. A relevant characteristic about these models is their relation with the Veneziano-Yankielowicz system \cite{Veneziano1982}, which is a low energy effective model for supersymmetric QCD. The parameter $N$ is connected with the number of colors in the $SU(N)$ gauge group. It is well known that the bosonic sector of these theories involves the presence of topological defects connecting the different vacua of the model. In (3+1)-dimensions these solutions arise as \textit{domain walls}, smooth solutions for the field equations such that their energy density is a localized function in one of the directions whereas it has trivial dependence in the others. These static solutions have been extensively studied in generalized WZ models for different values of $N$, see, for example, \cite{Abraham1991,Cvetic1991,Dvali1999,Dvali1999b,Chivisov1997,Gabadadze2000} and references therein. It is clear that the dimensional reduction of the theory to the relevant direction turns \textit{domain walls} into \textit{kinks}. On the other hand, domain wall junctions have also been thoroughly analyzed for these WZ models in two dimensions, see \cite{Saffin1999,Bazeia2000,Bazeia2000b,Binosi2000,Carroll2000,Gibbons1999}. As a remark, the majority of the previously cited papers focus on the case $N=3$, where the model exhibits a three-fold rotational symmetry $\mathbb{Z}_3$. This means that, for the effective $(1+1)$-dimensional complex field theory models derived from these WZ models, identical kinks carrying three different colors emerge. This color is an intrinsic property of the kinks, which is determined by the topological sector where they live.

From a more general point of view, this kind of models can be generalized to the case where the model has no specific symmetry by introducing a superpotential $W(\phi)$ determined by a general $N+1$-degree polynomial in $\phi$, such that $W'(\phi)$ presents $N$ arbitrary simple roots. An interesting range of possibilities arise in this situation because, depending on their relative positions in the complex plane, not all the pairs of vacua are connected by a topological defect. This kind of generalizations has been constructed from different perspectives, see \cite{Cecotti1993,Abraham1991,Vafa1990,Bazeia2009,Abel2019}. In this paper we shall deal with a general $N=3$ case with three arbitrary vacua. With an adequate affine transformation in the complex plane, the general study of this model can be reduced to the case of two fixed vacua and only an arbitrary one, whose location is set by the model parameter \cite{Bazeia2009}. Cases evolving from three collinear vacua to three vacua located at the vertices of an equilateral triangle will be investigated. The previously mentioned vacuum arrangement affects the structure of the kink variety, giving rise to two different regimes. In the first one, two equivalent BPS kinks/antikinks (that we will call blue and red) arise whereas in the second one a new different BPS kink/antikink (called green) emerges. There is a special value of the parameter where the model has a three-fold rotational symmetry. In this case the three topological defects are identical. The goal of this work is to study the kink scattering in this family of models.

Kink scattering is a very interesting issue by itself. The discovery of a fractal structure in the final versus initial velocity diagram in this type of events for the $\phi^4$ model drove an intense study of the collision between these topological defects in non-integrable models. The scattering between kinks and antikinks in the $\phi^4$ model was initially addressed in the seminal references \cite{Sugiyama1979, Campbell1983, Anninos1991, Kudryavtsev1975}. In this case, there are only two different scattering channels: \textit{bion formation} and \textit{kink reflection}. In the first type of events, kink and antikink collide and bounce back over and over forming a \textit{bion} whereas in the second type, kink and antikink emerge after the impact and move away with a certain final velocity $v_f$. The bion formation regime arises for low collision velocities $v_0$ while the kink reflection regime emerges for large values of $v_0$. The transition between these two regimes shapes a fractal pattern where the two previously mentioned scattering channels are interlaced. Besides, the kink reflection windows arising in this range are associated with $n$-bounce scattering events where kink and antikink collide several times before definitely escaping. The presence of these $n$-bounce windows can be explained by means of the \textit{resonant energy transfer mechanism}, which allows an energy exchange between the zero and shape kink modes at every impact. In these events, kink and antikink repeatedly collide and bounce back until the energy redistribution favors enough the zero mode to allow kink and antikink to escape and move away. Note that the scattering between wobbling kinks in this theory has been considered in \cite{Alonso2021}. The $\phi^4$ model is only the first example with these properties. The resonant energy transfer mechanism and other related phenomena are present in a large variety of one-component scalar field theory models, such as the double sine-Gordon model \cite{Shiefman1979, Peyrard1983, Campbell1986, Gani1999, Malomed1989, Gani2018, Gani2019}, deformed $\phi^4$ models \cite{Simas2016,Gomes2018,Bazeia2017b,Bazeia2017a, Bazeia2019, Adam2019, Romanczukiewicz2018, Adam2020, Mohammadi2020, Yan2020, Dorey2018}, $\phi^6$-models \cite{Romanczukiewicz2017, Weigel2014, Gani2014, Bazeia2018b, Lima2019, Marjaheh2017, Dorey2011} and other more complex models \cite{Mendoca2019, Belendryasova2019, Zhong2020, Bazeia2020c, Christov2019, Christov2019b, Christov2020,Campos2020}. In general, the number of scattering channels arising in one-component scalar field theory models is reduced. For example, for the $\phi^4$ model there exist only two types of scattering events. The situation is completely different for two-component models, where exotic scattering processes can take place due to the fact that the internal space goes from having one dimension to having two dimensions. The kink dynamics in some models of this class has been analyzed, for example, in references \cite{Halavanau2012, Romanczukiewicz2008, Alonso2018, Alonso2018b,Alonso2017, Alonso2019, Ferreira2019}. The previously mentioned \textit{kink reflection} and \textit{bion formation} processes are also found in these models, but \textit{kink-antikink annihilation}, \textit{kink-antikink transmutation}, \textit{kink reflection} (where internal charges are exchanged) are some of the novel scattering channels that have been identified in this framework. Furthermore, new scattering scenarios can be considered. For instance, linearly stable non-topological kinks arise in some two-component scalar field theory models, which carry a winding charge \cite{Alonso2020}. The collision between two kinks with the same winding charge can give place to exotic scattering events where a half of each kink annihilate each other and the other two halves recombine to form one new non-topological kink with the reversed winding charge. There are other initial velocity windows where kinks reflect each other but their winding charges are reversed by the impact.

In this paper the scattering between the two symmetric (red and blue) kinks arising in the previously commented model is thoroughly analyzed. Two different scattering channels are found: \textit{kink-kink reflection} and \textit{kink-kink hybridization}. In the last type of events, the collision between the two (blue and red) symmetric kinks gives rise to the third different (green) kink. For some initial velocity windows the resonant energy transfer mechanism is activated allowing a (green) vibrating kink to split into two symmetrical kinks moving away.

The organization of this paper is as follows: in Section \ref{sec2} the model is introduced and the static BPS kinks are described. The linear stability study of these solutions is also addressed. Section \ref{sec3} is devoted to the kink scattering analysis in this model. A classification of the possible scattering scenarios is provided and the scattering events associated with the collision between the two symmetric kinks are discussed. Some conclusions are drawn in Section \ref{sec4}. Finally, the numerical method designed to carried out the kink scattering simulations in this model is thoroughly described in the Appendix.


\section{The model} \label{sec2}

The dimensional reduction of the bosonic sector of the generalized (3+1)-dimensional ${\cal N}=2$ supersymmetric Wess-Zumino model leads to an effective (1+1)-dimensional complex scalar field theory model whose action functional reads as
\begin{equation}\label{complexaction}
S=\int d^2 x \Big[ \frac{1}{2} \partial_\mu \phi \, \partial^\mu \overline{\phi} - \frac{1}{2} \, \frac{\partial W}{\partial \phi} \, \overline{\frac{\partial W}{\partial \phi}} \Big]  \hspace{0.4cm},
\end{equation}
where the superpotential $W$ is a holomorphic function $W=W(\phi)$. Here, $\phi:\mathbb{R}^{1,1} \rightarrow \mathbb{C}$ is a complex scalar field and overlines stand for complex conjugation. Einstein summation convention is used in (\ref{complexaction}) for $\mu=0,1$ and fields and spacetime coordinates $x^0\equiv t$ and $x^1\equiv x$ are assumed to be dimensionless. The Minkowski metric $g_{\mu\nu}$ has been chosen as $g_{00}=-g_{11} =1$ and $g_{12}=g_{21}=0$. In general, nonlinear terms will be introduced by the potential density (independent of the spatial derivatives) given by the non-negative expression
\begin{equation} \label{complexpotential}
U(\phi, \overline{\phi}) = \frac{1}{2} \, \frac{\partial W}{\partial \phi} \, \overline{\frac{\partial W}{\partial \phi}}  \hspace{0.4cm}.
\end{equation}
Note that the rotation of the superpotential by an angle $\alpha$ in the complex internal plane
\begin{equation}\label{rotatedsuperpotential}
W^{(\alpha)}(\phi)= e^{i\alpha} W(\phi)
\end{equation}
with $\alpha \in [0,2\pi)$ provides us with the same potential term $U(\phi)$. In other words, there exists a one-parametric family of superpotentials describing the same physical system. The supersymmetric structure of the model underlies the presence of first order differential equations
\begin{equation} \label{complexedo}
\frac{\partial \phi}{\partial x} = \overline{\frac{\partial W^{(\alpha)}}{\partial \phi}} \hspace{0.5cm}, \hspace{0.5cm}
\frac{\partial \overline{\phi}}{\partial x} = \frac{\partial W^{(\alpha)}}{\partial \phi} \hspace{0.5cm}
\end{equation}
which characterize the BPS solutions of the model. They represent static solutions for the field equations
\begin{equation} \label{complexedp}
\frac{\partial^2 \overline{\phi}}{\partial t^2} - \frac{\partial^2 \overline{\phi}}{\partial x^2} + \frac{\partial^2 W}{\partial \phi^2} \overline{\frac{\partial W}{\partial \phi}} = 0  \hspace{0.5cm}, \hspace{0.5cm}  \frac{\partial^2 \phi}{\partial t^2} - \frac{\partial^2 \phi}{\partial x^2}+ \overline{\frac{\partial^2 W}{\partial \phi^2}} \frac{\partial W}{\partial \phi} = 0
\end{equation}
that minimize the energy functional
\begin{equation}\label{complexenergy}
E=\int d x \  \varepsilon (x,t) = \int d x \Big[ \frac{1}{2} \frac{\partial \phi}{\partial t} \, \frac{\partial \overline{\phi}}{\partial t} + \frac{1}{2} \frac{\partial \phi}{\partial x} \, \frac{\partial \overline{\phi}}{\partial x} + \frac{1}{2} \, \frac{\partial W}{\partial \phi} \, \overline{\frac{\partial W}{\partial \phi}} \Big]
\end{equation}
in a specific topological sector. For static configurations $\phi=\phi(x)$, (\ref{complexenergy}) can be written a la Bogomol'nyi:
\begin{equation}\label{complexenergyBogo}
E= \frac{1}{2} \int dx \left| \frac{d \phi}{dx} - \overline{\frac{\partial W^{(\alpha)}}{\partial \phi}} \right|^2 + \frac{1}{2} \left| \int d ( W^{(\alpha)}+\overline{W^{(\alpha)}}) \right| \hspace{0.5cm}.
\end{equation}
An essential role among these BPS solutions is played by the topological defects or kinks. Kinks are finite energy solutions of the field equations (\ref{complexedp}) whose energy densities are localized. This implies that these solutions can be interpreted as extended particles of the physical system.

The holomorphy of the superpotential can be used to easily identify the orbits of these BPS solutions \cite{Alonso2000,Bazeia2009,Bazeia2008}. BPS static kinks satisfy first-order ordinary differential equations (\ref{complexedo}), in such a way that
\begin{equation} \label{dexactsupepotential}
\frac{\partial W^{(\alpha)}}{\partial \phi} d \phi - \overline{\frac{\partial W^{(\alpha)}}{\partial \phi}} d \overline{\phi} =0 \quad , \quad  d ( W^{(\alpha)}(\phi)- \overline{ W^{(\alpha)}(\phi)}) =0 \hspace{0.5cm},
\end{equation}
which implies the simple relation
\begin{equation} \label{kinkorbits}
{\rm Im} W^{(\alpha)}(\phi) = \gamma \hspace{0.5cm},
\end{equation}
for the orbits of the solutions in the complex internal space, and where $\gamma$ is a real constant. Kink orbits are horizontal straight lines in the $W^{(\alpha)}$-plane, or, equivalently, straight lines in the $W$-plane. From now on we shall focus on a family of models, whose superpotential (\ref{rotatedsuperpotential}) is given by the expression
\begin{equation} \label{deltasuperpotential}
W^{(\alpha)} (\phi)=  e^{i \alpha}  \Big[\delta \phi \ +\ (\delta - 1) \Big( \frac{1}{2} \phi^2+\frac{1}{3} \phi^3 \Big) \ - \frac{1}{4} \phi^4 \Big] \hspace{0.5cm} .
\end{equation}
The associated potential (\ref{complexpotential}) is written as
\begin{equation}\label{potential}
U(\phi,\overline{\phi})= \frac{1}{2} \Big| (\phi^2 + \phi +1)(\delta -\phi) \Big|^2 \hspace{0.5cm} ,
\end{equation}
i.e., $U(\phi,\overline{\phi})$ presents three zeroes (absolute minima). The set of vacuum solutions ${\cal M}=\{\phi \in \mathbb{C}: U(\phi,\overline{\phi})=0\}$ is formed by these three zeroes:
\begin{equation}\label{vacua}
{\cal M} = \Big\{ p_0=\delta  \, , \, p_1= {\textstyle -\frac{1}{2} + i \frac{\sqrt{3}}{2}  } \, , \, p_2= {\textstyle -\frac{1}{2} -i \frac{\sqrt{3}}{2}  }  \Big\} \hspace{0.5cm} ,
\end{equation}
two of them are fixed, $p_1$ and $p_2$, whereas the other one, $p_0$, can move in ${\mathbb C}$ for the different values of the $\delta$ parameter. It has to be remarked that any other model of this type with three different vacua can be reduced to this one with an appropriate affine transformation in ${\mathbb C}$.

\begin{figure}[h]
\centerline{\includegraphics[height=5cm]{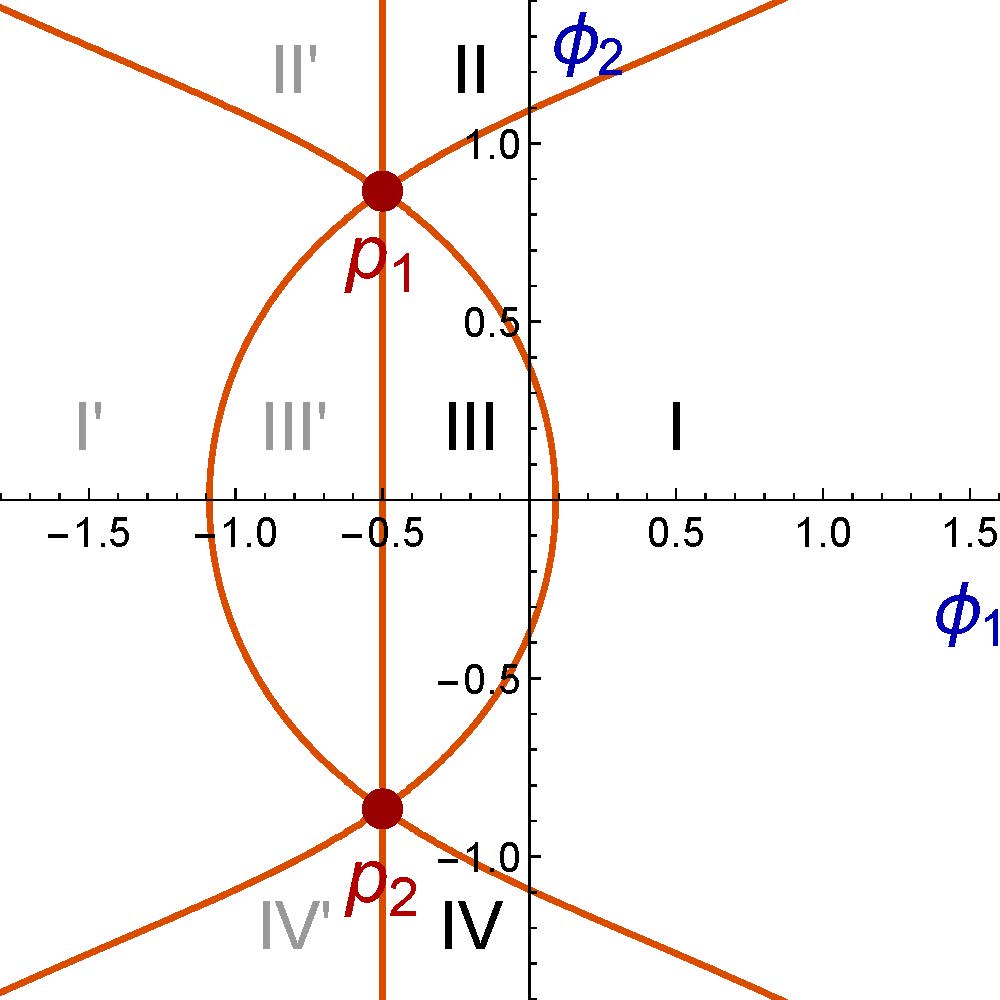} }
\caption{\small Graphics of the marginal stability curve (\ref{marginalcomplex}).}\label{fig:marginal}
\end{figure}

The finite energy configuration space ${\cal C}$ is the union of nine topologically disconnected sectors, each of them characterized by the pair of elements of ${\cal M}$ which are asymptotically connected by the elements of ${\cal C}$. Therefore, ${\cal C}=\sqcup_{i,j=0,1,2} \, {\cal C}_{ij}$ where ${\cal C}_{ij}$ stands for the set of configurations which comply with
\begin{equation} \label{asymptotic}
\lim_{x\rightarrow -\infty} \phi(x)=p_i \hspace{0.5cm}\mbox{and}\hspace{0.5cm} \lim_{x\rightarrow \infty} \phi(x)=p_j
\end{equation}
Topological kinks belong to some sector ${\cal C}_{ij}$ with $i\neq j$ while non-topological kinks are included in some ${\cal C}_{ii}$. As previously mentioned, BPS kink orbits are determined by the implicit condition (\ref{kinkorbits}) for some real constant $\gamma$ and angle $\alpha$ to be calculated in such a way that asymptotic conditions (\ref{asymptotic}) are satisfied, i.e. condition
\begin{equation}\label{alfa}
{\rm Im} \, W^{(\alpha)}(p_i) = {\rm Im} \, W^{(\alpha)}(p_j)
\end{equation}
provides us the values $\alpha = \alpha^{(ij)}$ (mod $\pi$) such that equation (\ref{kinkorbits}) describes the orbits belonging to ${\cal C}_{ij}$, or ${\cal C}_{ji}$ because: $\alpha^{(ij)}=\alpha^{(ji)} + \pi$, and $\gamma$ is determined by the value of (\ref{alfa}) for $\alpha = \alpha^{(ij)}$ in each topological sector. However, the existence of values $\alpha^{(ij)}$ complying with (\ref{alfa}) does not guarantee the existence of an element in ${\cal C}_{ij}$, because kink solutions only occur if any of trajectories defined by (\ref{kinkorbits}), for $\alpha = \alpha^{(ij)}$, asymptotically connects $p_i$ and $p_j$ by means of a finite arc in ${\mathbb C}$. This is a delicate subject, because as it is well known \cite{Cecotti1993,Abraham1991,Vafa1990,Bazeia2009}, the algebraic curve (\ref{kinkorbits}) can contain, or not, this kind of finite arcs in ${\mathbb C}$. It can be proved that in this family of models there exist different regions determined by the $\delta$-parameter where the number of kinks is different. Taking into account that the value of the complex parameter $\delta$ and the vacuum $p_0$ coincide, the previously mentioned regions can be visualized in the $\phi_1$-$\phi_2$ plane as locations of the vacuum $p_0$ for which different number of topological defect solutions can be found. The boundaries between these regions are determined by the so called \textit{marginal stability curve}. Defining the real and imaginary parts of $\delta$ as: $\delta = \delta_1+i\delta_2$, this curve is given by the alignment condition for the three vacua in the complex $W$-plane, and results to be:
\begin{equation}
\label{marginalcomplex}
(2\delta_1+1) \left(2\delta_1^4-6\delta_2^4-4\delta_1^2\delta_2^2+4\delta_1^3-4\delta_1 \delta_2^2+12 \delta_1^2+8\delta_2^2+10 \delta_1-1\right)=0
\end{equation}
for our model. This curve is represented in the $\phi=\phi_1+i\phi_2$ plane in Figure \ref{fig:marginal}, where the two fixed vacua of ${\cal M}$ are showed and any choice of $\phi=\delta$ in the plane establish a relative position between the three vacua.

There appear eight different regions with a mirror symmetry with respect to the line $\delta_1=-\frac{1}{2}$. Thus cases I, II, III and IV are equivalent to I', II', III' and IV' respectively, and the problem is reduced to only four possible situations: Region I stands for the existence of three orbits in (\ref{kinkorbits}) connecting all the three possible pairs of vacua in ${\cal M}$, whereas regions II, III and IV represent the three possible cases of absence of one of these orbits. We conclude, from a qualitative point of view, that there exits only two relevant situations: first, the case where all the vacua are connected by a kink orbit (region I, or equivalently I'), and, secondly, only two kink orbits are present and there are not exist kinks between two of the vacua in ${\cal M}$ (regions II, III and IV and their respective equivalent regions).

For the sake of simplicity, we shall study the kink scattering in this model when $\delta$ is a real parameter, that is, $\delta_2=0$. As a consequence the vacuum $p_0$ moves along the axis $\phi_1$ as the value of the parameter $\delta= \delta_1$ varies. Now, equation (\ref{marginalcomplex}) reduces to:
\begin{equation} \label{stabilitycurve}
(2\delta+1)(2\delta^4+4\delta^3+12\delta^2+10\delta-1)=0
\end{equation}
and the intervals in ${\mathbb R}$ which characterize the presence of a different number of kinks are determined by the three roots of (\ref{stabilitycurve})
\begin{equation} \label{criticalpoints}
\delta_a = -\frac{1}{2}-\frac{1}{2}\sqrt{3(2\sqrt{3}-3)} \hspace{0.3cm} , \hspace{0.3cm}  \delta_b = -\frac{1}{2} \hspace{0.3cm} , \hspace{0.3cm} \delta_c=-\frac{1}{2}+\frac{1}{2}\sqrt{3(2\sqrt{3}-3)} \hspace{0.3cm}.
\end{equation}
For $\delta < \delta_a$, there exist kinks connecting the three possible pairs of vacua; for $\delta_a < \delta < \delta_c$, there are only two possible kinks because $p_1$ and $p_2$ are not connected by a finite arc of (\ref{kinkorbits}) and for $\delta > \delta_c$ there are again three kinks joining the vacua. Indeed, the above commented mirror symmetry with respect to the value $\delta_b=-\frac{1}{2}$, which reduces the significant cases to the parameter range $\delta \in [-\frac{1}{2},\infty]$. In this paper we shall explore the kink scattering properties in the range $\delta\in [-0.5,1]$, which are representative of the global model.

\begin{figure}[h]
\centerline{\includegraphics[height=3.5cm]{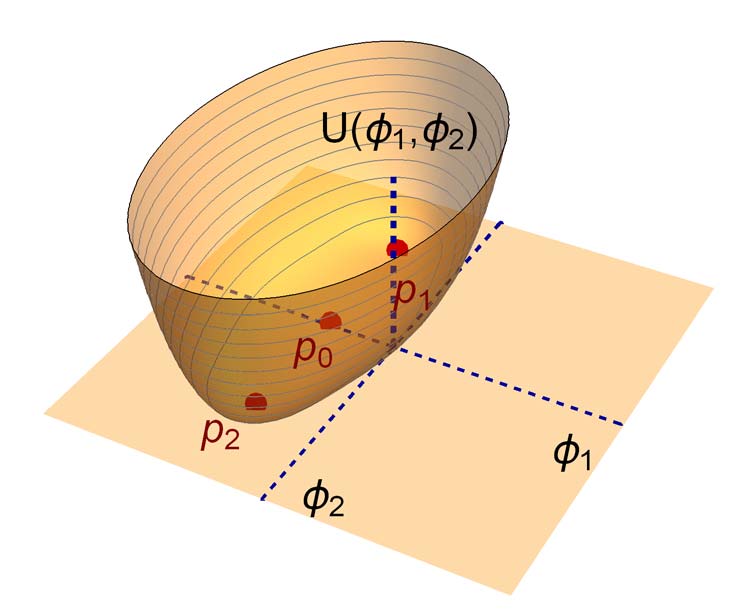} \hspace{0.05cm} \includegraphics[height=3.5cm]{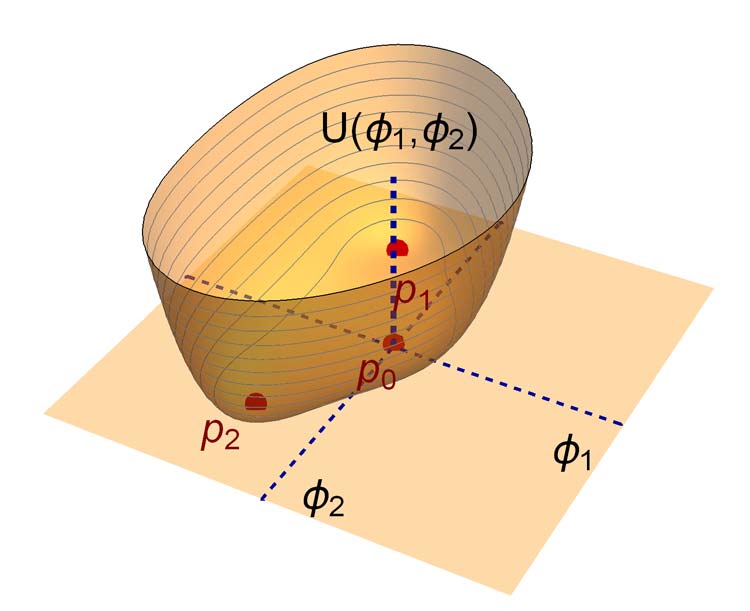} \hspace{0.05cm} \includegraphics[height=3.5cm]{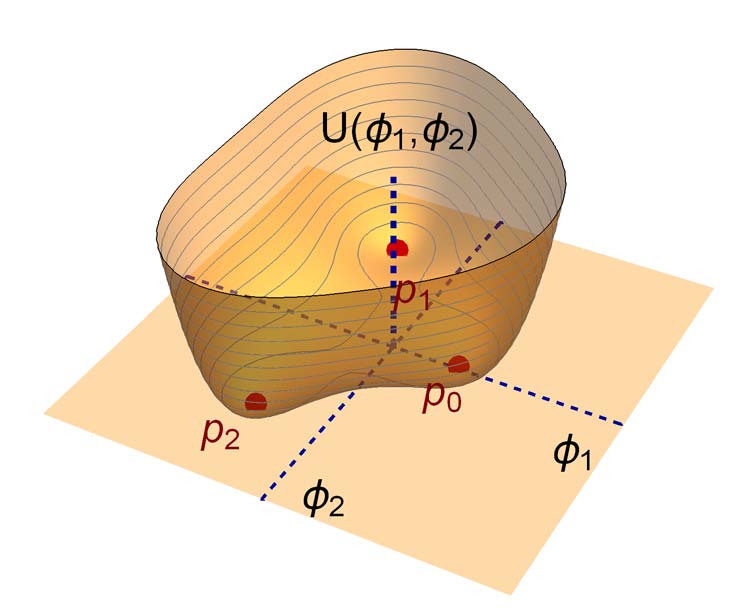} \hspace{0.05cm} \includegraphics[height=3.5cm]{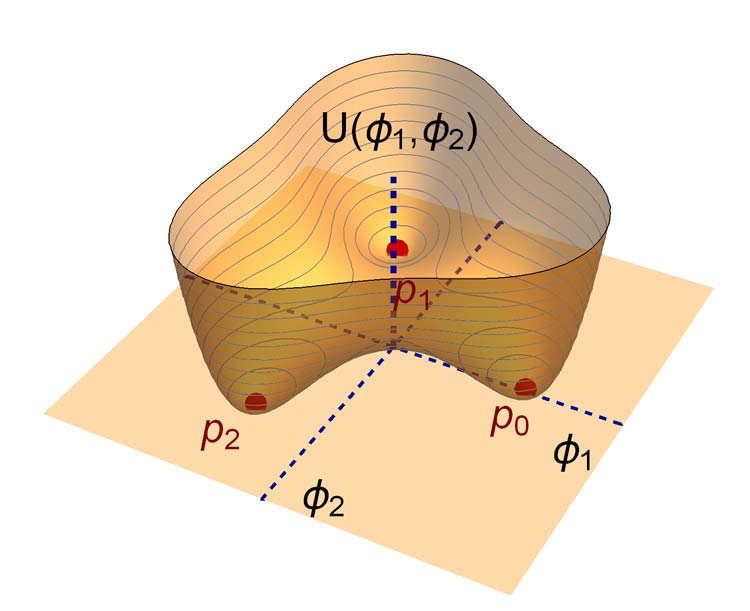}}
\caption{\small Graphics of the potential term $U(\phi,\overline{\phi})\equiv U(\phi_1,\phi_2)$ for the parameter values: (a) $\delta=-1/2$, (b) $\delta=0$, (c) $\delta=1/2$, (d) $\delta=1$. }\label{fig:potential}
\end{figure}

For $\delta=-\frac{1}{2}$, the potential (\ref{potential}) can be expressed in components as
\[
U(\varphi_1,\varphi_2) =\frac{1}{2} (\varphi_1^2 + \varphi_2^2) \Big[(\varphi_1^2 + \varphi_2^2)^2 + \frac{3}{2} (\varphi_1^2 - \varphi_2^2) + \frac{9}{16} \Big]
\]
where the new fields $\varphi_1=\phi_1-\frac{1}{2}$ and $\varphi_2=\phi_2$ have been employed. This redefinition allows us to observe that $U(0,\varphi_2)=\frac{1}{2} \varphi_2^2 (\varphi_2^2 - \frac{3}{4} )^2$, that is, the one-component $\varphi_2^6$ model is embedded in our model. This means that the three vacua are aligned in the complex plane, see Figure \ref{fig:potential}(a). Another significant situation arises when $\delta=1$ and the potential (\ref{potential})
\[
U(\phi_1,\phi_2) =\frac{1}{2} \Big[(\phi_1^2 + \phi_2^2)^3 - 2\phi_1 (\phi_1^2 - 3\phi_2^2) + 1 \Big]
\]
exhibits a three-fold rotational symmetry ${\mathbb Z}_3$, see Figure \ref{fig:potential}(d). In this case the vacua are located at the vertices of an equilateral triangle. Any other choice of $\delta\in {\mathbb R}$ determines a structure of isosceles triangle for the elements of ${\cal M}$, with an obvious ${\mathbb Z}_2$ symmetry associated to reflections across the $\phi_2$-axis. Figure \ref{fig:potential} displays a sequence of graphics of the potential (\ref{potential}) and the vacuum arrangements for several values of $\delta$.

For the sake of clarity, the topological charge defined by
\begin{equation} \label{topologicalcharge}
Q= \frac{3}{2\pi} \Big( {\rm arg} \, \phi(+\infty) - {\rm arg} \, \phi(-\infty) \Big) \,{\rm mod}\,3 =(i-j)\,{\rm mod}\,3  \,\equiv \, [i-j]
\end{equation}
will be employed to classify the kink variety. As it is well known, if $K(x)$ represents a kink solution of our model the mirror image of this solution $K(-x)$ is also a solution, which is referred to as \textit{antikink}. For kinks living in two-dimensional internal spaces a convention must be chosen to distinguish between kinks and antikinks. The charge (\ref{topologicalcharge}) can fulfill this task. The value $Q=1$ will characterize \textit{kinks} whereas $Q=-1$ will be associated with \textit{antikinks}. Non-topological kinks involve a charge $Q=0$.

\begin{figure}[h]
\centering\begin{tabular}{cc}\includegraphics[height=4.cm]{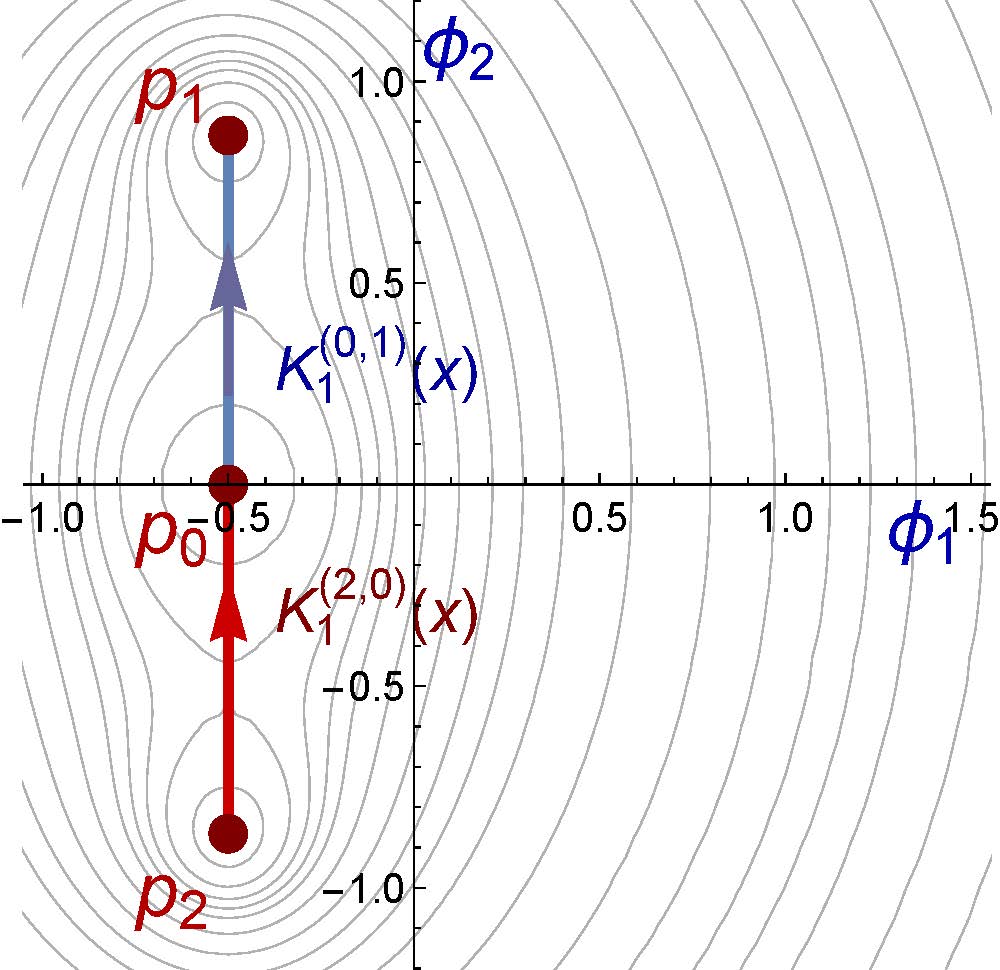} \hspace{0.5cm} & \hspace{0.5cm} \includegraphics[height=4.cm]{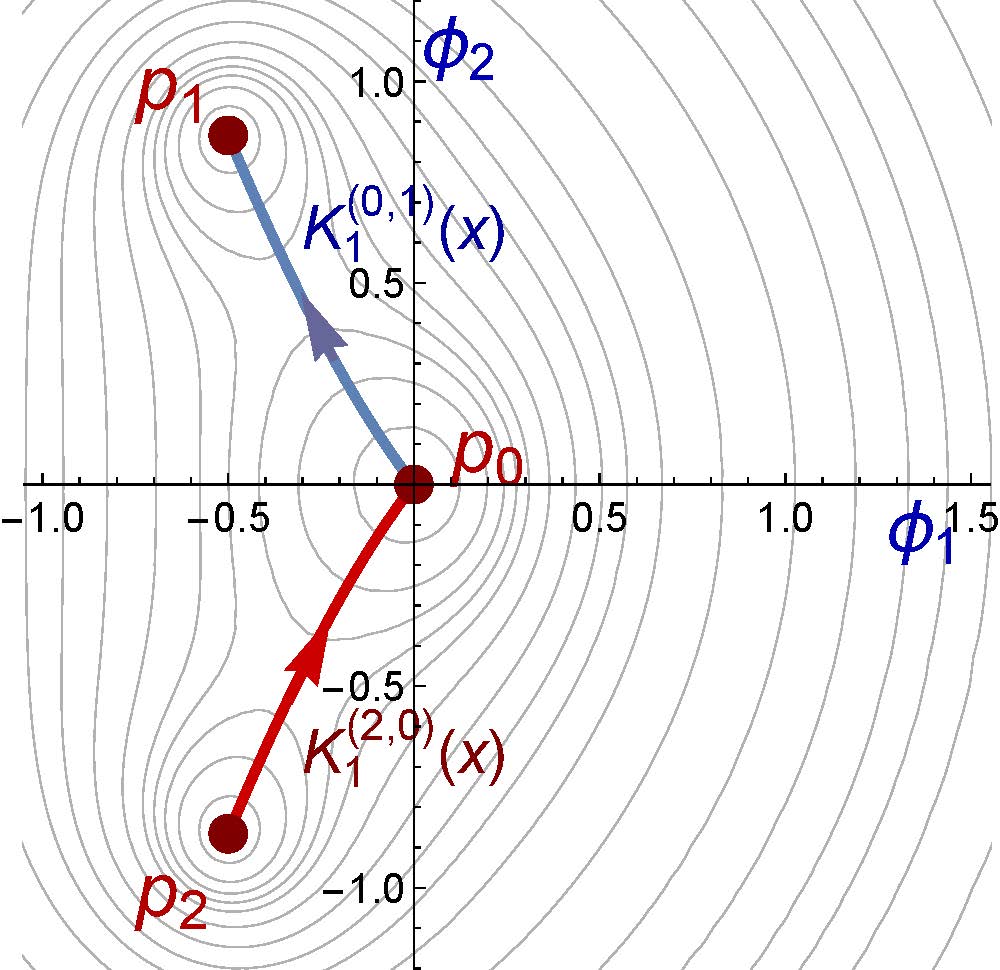} \\ & \\  \includegraphics[height=4.cm]{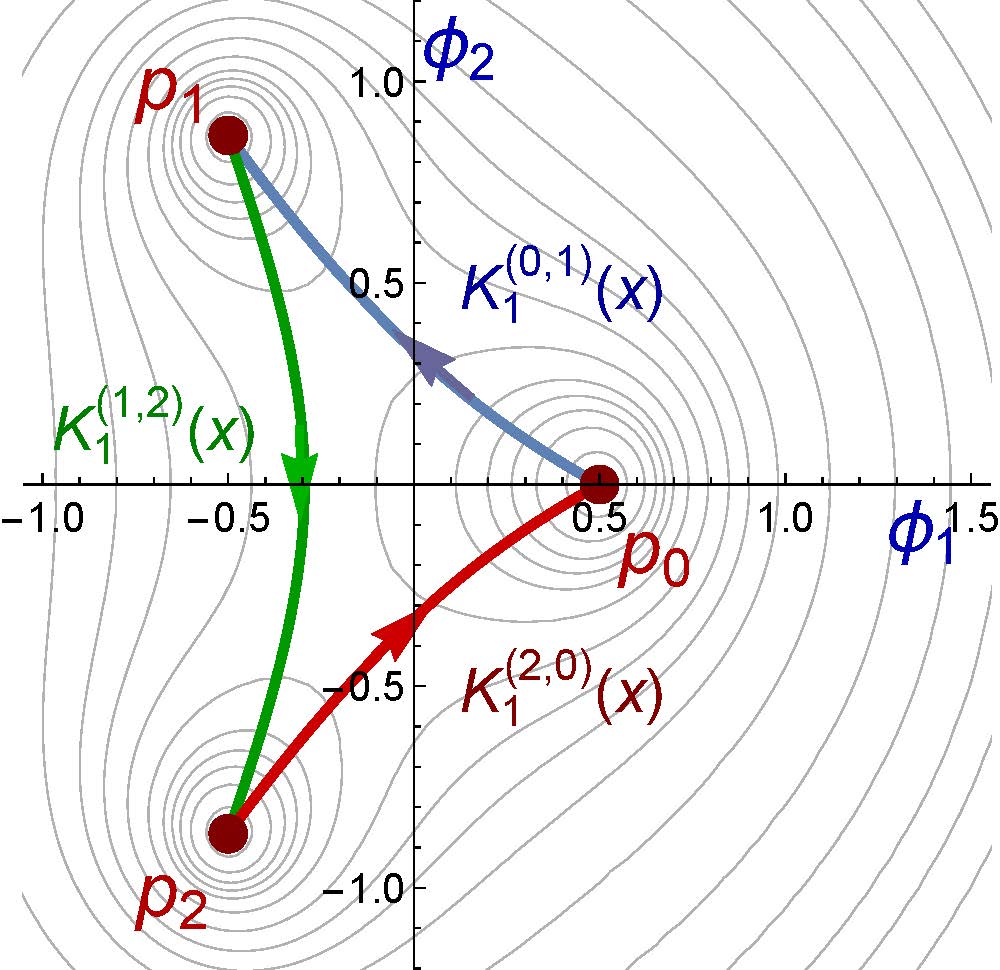} \hspace{0.5cm} & \hspace{0.5cm} \includegraphics[height=4.cm]{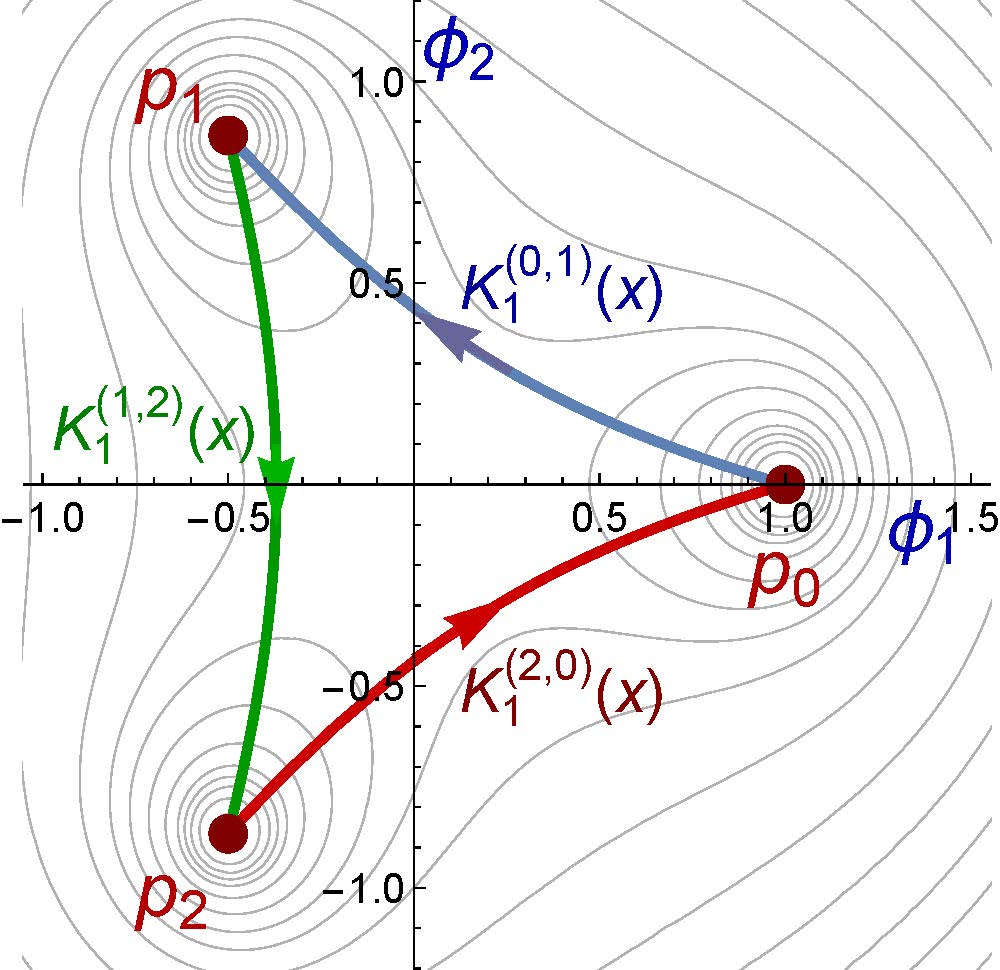} \end{tabular}
\caption{\small Graphical representation of the orbits described by the static BPS kinks for the parameter values: (a) $\delta=-1/2$, (b) $\delta= 0$, (c) $\delta= 1/2$ and (d) $\delta=1$. The potential $U(\phi,\overline{\phi})$ is also represented by means of an underlying contour plot.}\label{fig:KinkOrbits1}
\end{figure}

It is thus adequate to introduce the notation $K_{Q}^{(i,j)}(x) $ to designate the static BPS kink solutions which verify (\ref{asymptotic}) with topological charge $Q$, connecting the vacua $v_i$ and $v_j$. Moreover, a color code will be used to simplify this explicit notation, see Table \ref{fig:KinkOrbits2} and Figure \ref{fig:KinkOrbits1}.

\begin{table}[h]
	\centering {\small\begin{tabular}{ll} \hline Kinks & \\ \hline\hline $K_1^{(0,1)} \equiv K_b$  & \hspace{0.2cm} blue kink  \\ $K_1^{(1,2)} \equiv K_g$  &\hspace{0.2cm} green kink  \\ $K_1^{(2,0)} \equiv K_r$  &\hspace{0.2cm} red kink \\\hline
	\end{tabular}} \hspace{0.3cm} {\small\begin{tabular}{ll} \hline
	Antikinks & \\ \hline\hline  $K_{-1}^{(1,0)} \equiv K_{\overline{b}}$  &\hspace{0.2cm} blue antikink \\  $K_{-1}^{(2,1)} \equiv K_{\overline{g}}$ & \hspace{0.2cm} green antikink \\  $K_{-1}^{(0,2)} \equiv K_{\overline{r}}$ & \hspace{0.2cm} red antikink \\\hline
	\end{tabular}}
	\caption{\small Color code to distinguish the intrinsic property of the kinks and antikinks.}\label{fig:KinkOrbits2}
\end{table}

For our model $K_b$ and $K_r$ kink orbits, associated respectively to the following solutions of equation (\ref{alfa}):
\begin{equation}
 \alpha^{(01)} = \arctan \Big[ \frac{-3\sqrt{3}(1+2\delta)}{(-1+10\delta +12\delta^2+ 4\delta^3+2\delta^4)}\Big]\  ,\quad
\alpha^{(20)}= \arctan \Big[ \frac{-3\sqrt{3}(1+2\delta)}{(1-10\delta -12\delta^2- 4\delta^3-2\delta^4)}\Big]\label{angleskinks}
\end{equation}
are present for any choice of $\delta \in {\mathbb R}$, and their antikinks $K_{\overline{b}}$ and $K_{\overline{g}}$ traverse the same orbit in the opposite direction, see Figure \ref{fig:KinkOrbits1}.  $K_g$ kink orbit, corresponding to the angle $\alpha^{(12)}= \frac{\pi}{2}$, does not exit if $\delta \in (\delta_a,\delta_c)$, see Figure \ref{fig:KinkOrbits1}(a) and (b), i.e. the algebraic equation (\ref{kinkorbits}), for $\alpha= \frac{\pi}{2}$, contains a finite arc connecting $p_1$ and $p_2$ only for $\delta < \delta_a$ or  $\delta> \delta_c$, see Figure \ref{fig:KinkOrbits1}(c) and (d).

\begin{figure}[h]
\centering\begin{tabular}{cccc}\includegraphics[height=2.5cm]{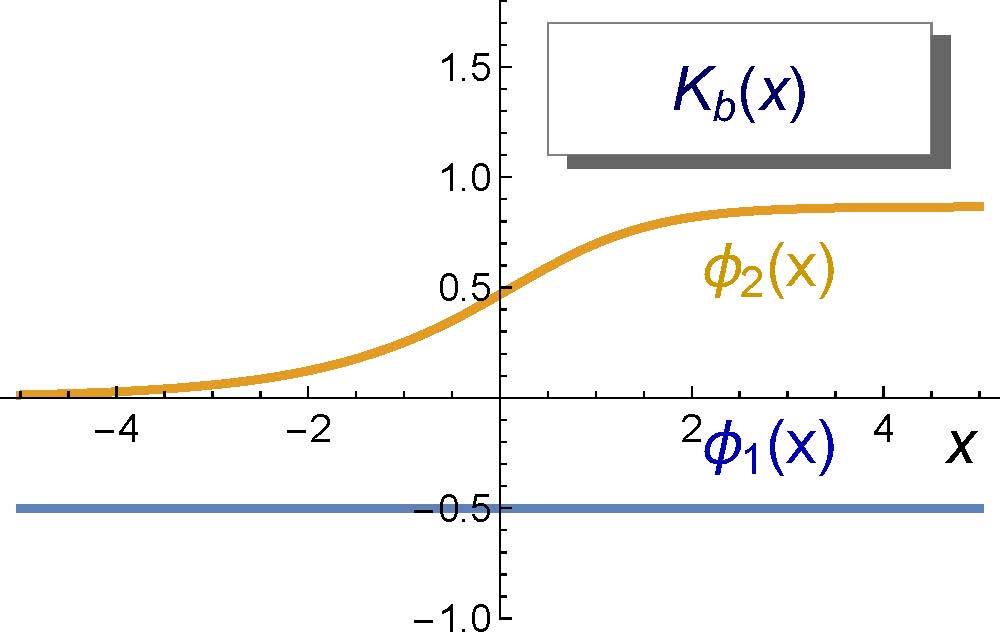} & \includegraphics[height=2.5cm]{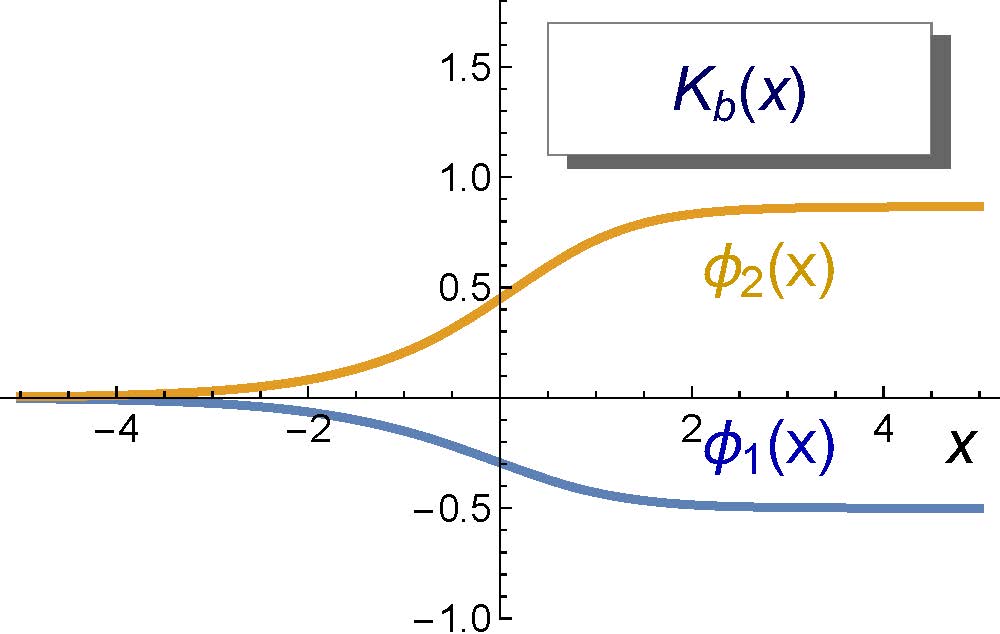} & \includegraphics[height=2.5cm]{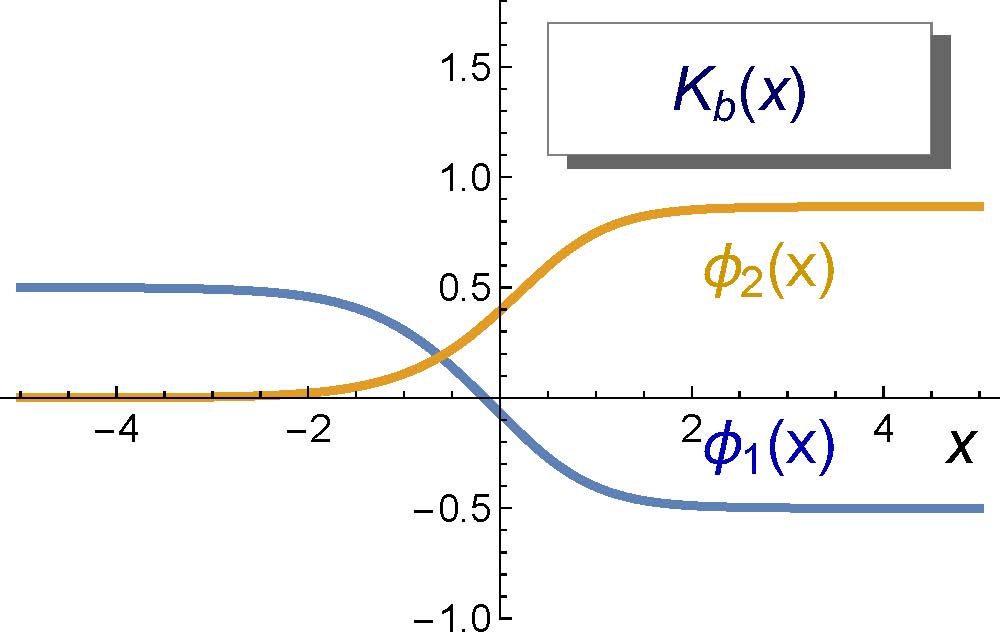} & \includegraphics[height=2.5cm]{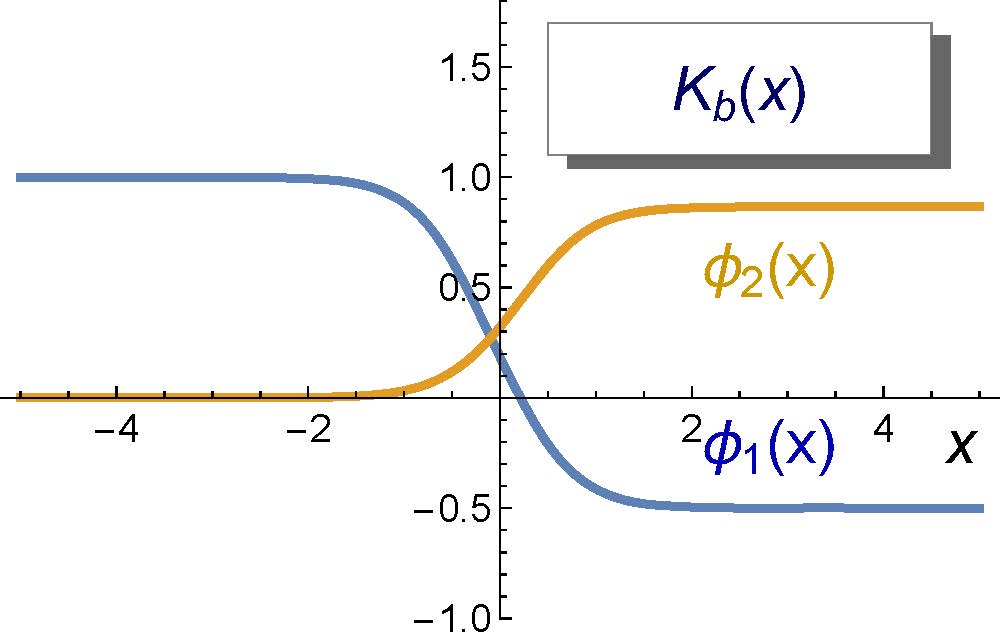} \end{tabular}
	\caption{\small Graphics of the profile of the static $K_b(x)$ kink with (a) $\delta=-1/2$, (b) $\delta=0$, (c) $\delta=1/2$ and (d) $\delta=1$.}
	\label{fig:Kinks1}
\end{figure}

\begin{figure}[h]
\centering\begin{tabular}{cccc}\includegraphics[height=2.5cm]{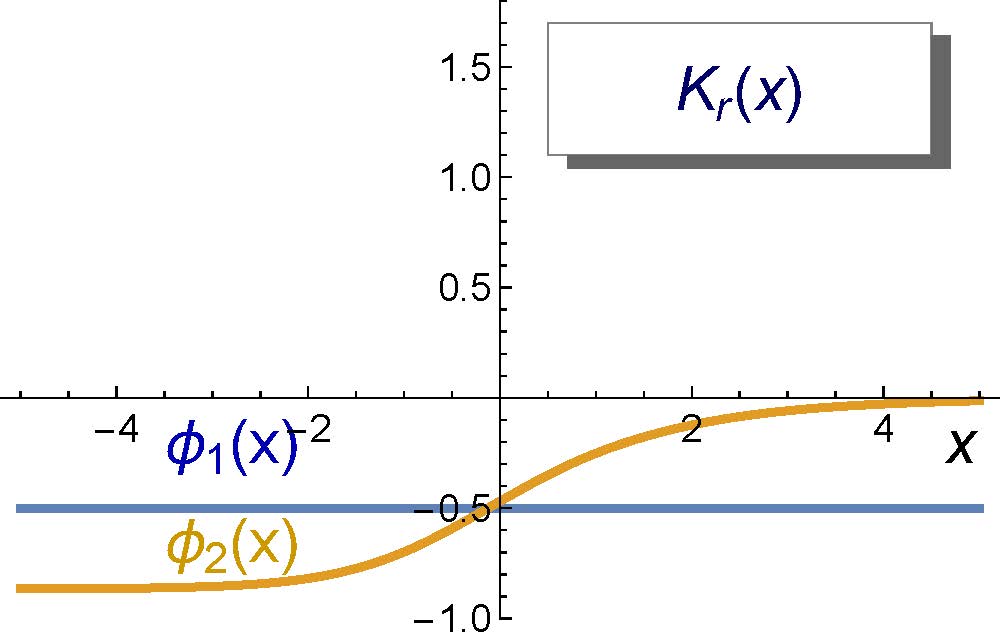} & \includegraphics[height=2.5cm]{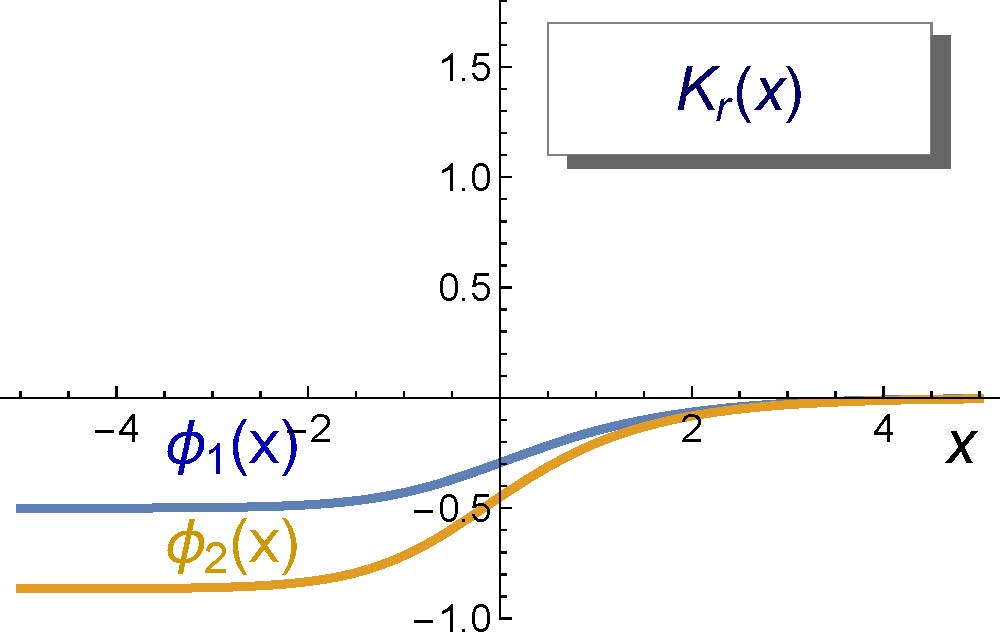} & \includegraphics[height=2.5cm]{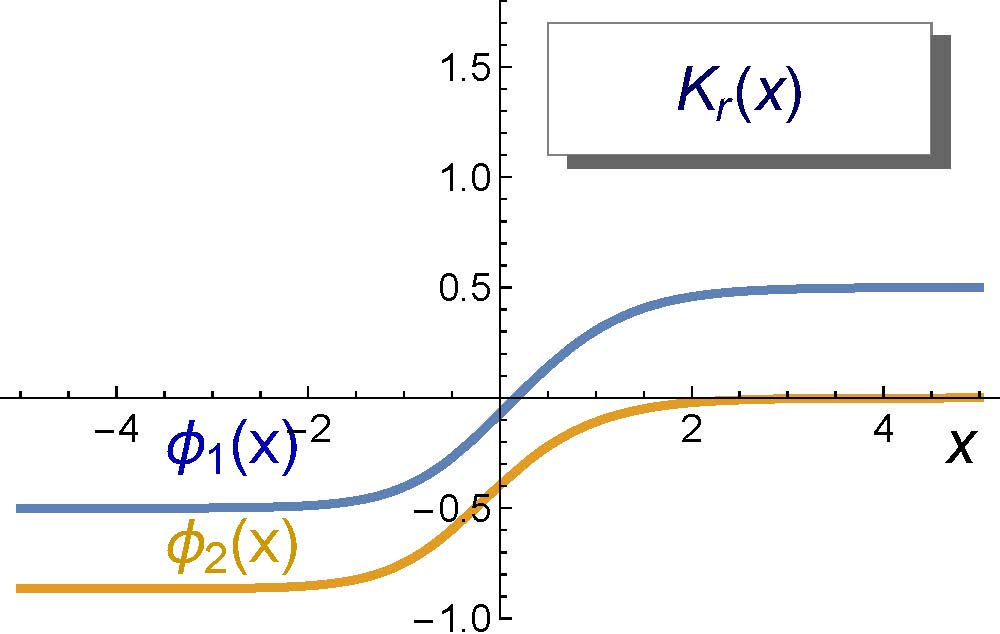} & \includegraphics[height=2.5cm]{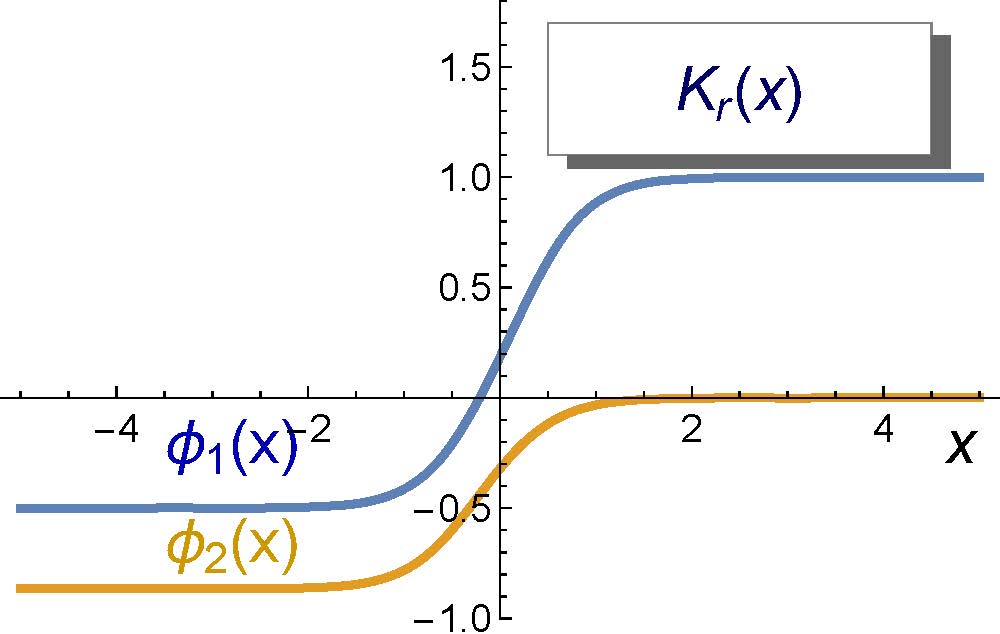} \end{tabular}
	\caption{\small Graphics of the profile of the static $K_r(x)$ kink with (a) $\delta=-1/2$, (b) $\delta=0$, (c) $\delta=1/2$ and (d) $\delta=1$.}
	\label{fig:Kinks2}
\end{figure}

\begin{figure}[h]
\centering\begin{tabular}{ccc}\includegraphics[height=2.5cm]{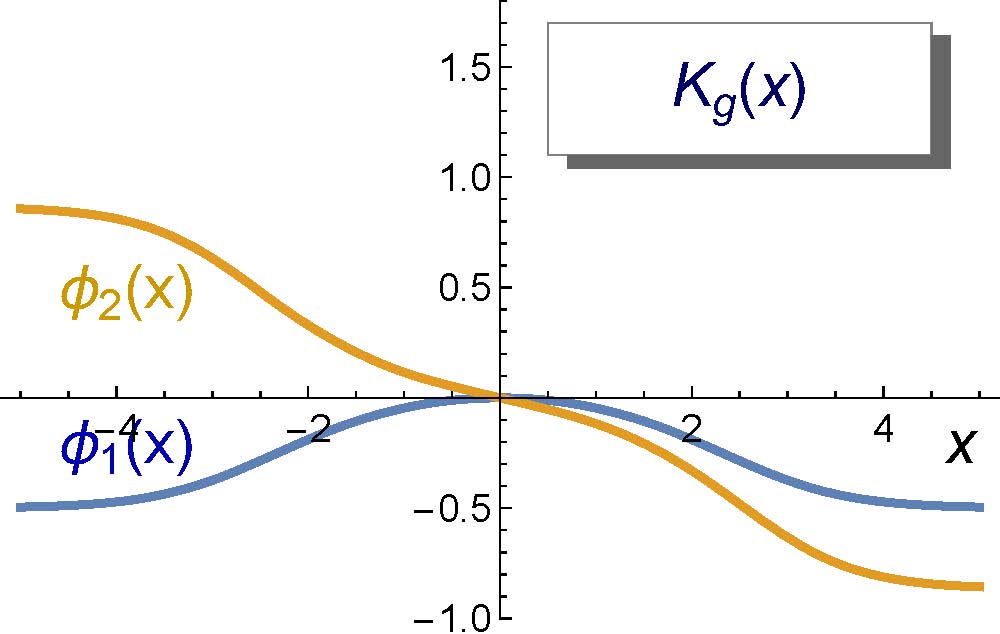} & \includegraphics[height=2.5cm]{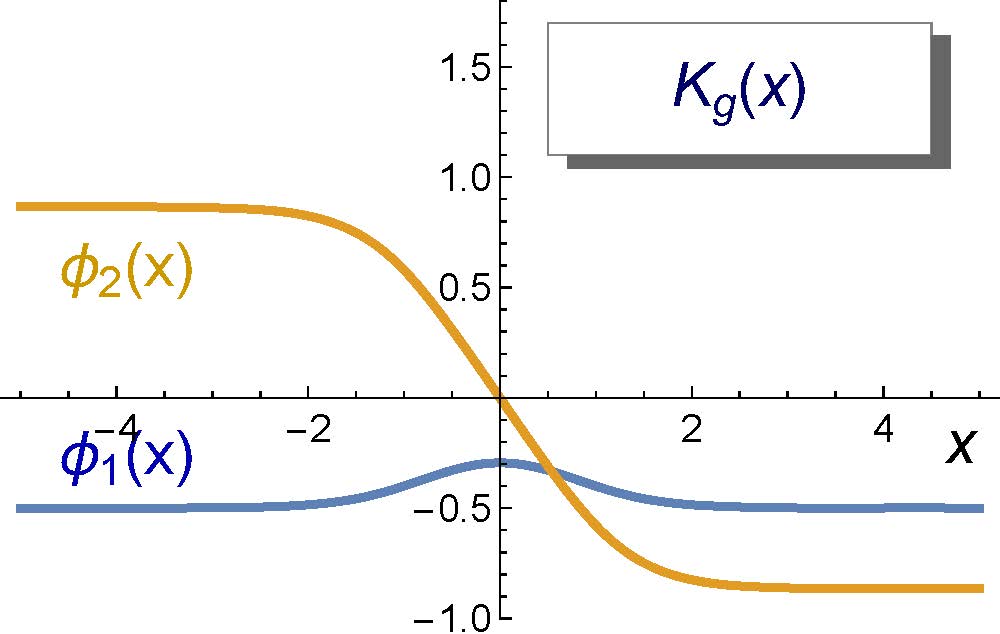} & \includegraphics[height=2.5cm]{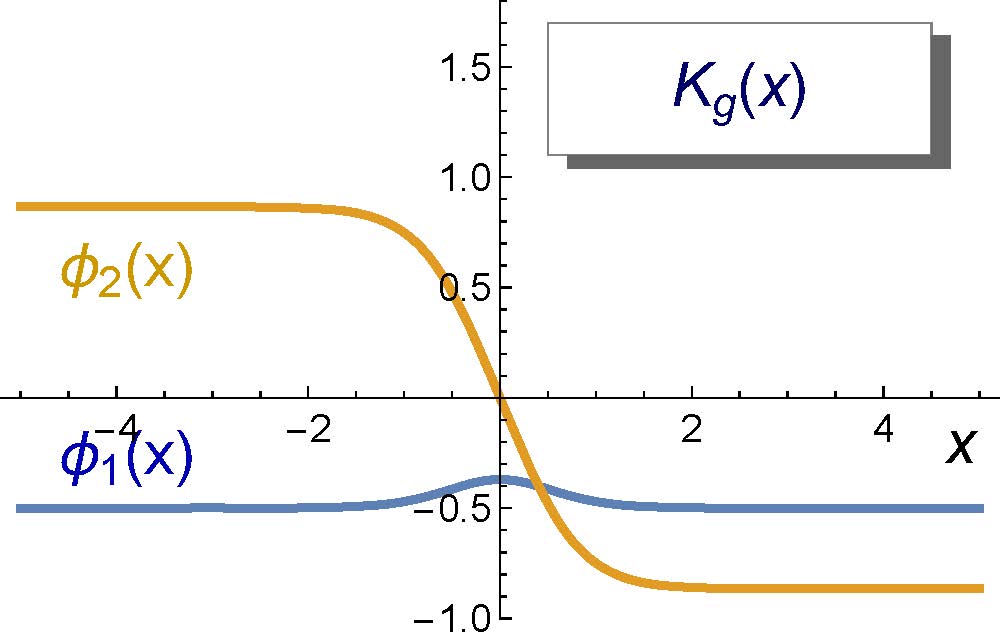}  \end{tabular}
	\caption{\small Graphics of the profile of the static $K_g(x)$-kink with (a) $\delta=0.1$, (b) $\delta=1/2$, (c) $\delta=1$.}
	\label{fig:Kinks3}
\end{figure}

Although the orbits (\ref{kinkorbits}) of the static BPS kinks have been analytically identified, it is not possible to extract an explicit expression of the kink profiles and numerical analysis must be applied to obtain their behaviors. The field components for the static blue $K_{b}(x)$, red $K_r(x)$ and green $K_{b}(x)$ kinks have been depicted in Figures \ref{fig:Kinks1}, \ref{fig:Kinks2} and \ref{fig:Kinks3} respectively for several values of $\delta$.

It can be checked that the kink energy is localized around a spatial point $x_C$. This is the reason behind the interpretation of kinks as extended particles. The point $x_C$ will be considered the center of these static BPS kinks. Notice that the field components of a kink valued at $x_C$ correspond with the midpoint of its trajectory, where the derivative of the solution and its distance to the vacua are maximized. It is clear that the kinks $K_b(x)$ and $K_r(x)$ (and their antikinks) are related by the reflection symmetry $\phi_2 \rightarrow  -\phi_2$. As a consequence, only two types of extended particle are described by the kinks and the antikinks arising in this model. The total energy of these static configurations, using (\ref{complexenergyBogo}), can be written as: $E[K^{(i,j)}(x)]=E[K^{(j,i)}(x)]= |W(p_j) - W(p_i)|$, and thus:
\begin{eqnarray}
&& E[K_b]=E[K_{\overline{b}}]=E[K_r]=E[K_{\overline{r}}]=\frac{1}{12} \sqrt{(1+\delta+\delta^2)^3 (7+\delta+\delta^2)} \label{energybr} \\
&& E[K_g]=E[K_{\overline{g}}]=\frac{\sqrt{3}}{4} (1+2\delta) \label{energyg}
\end{eqnarray}
Note that for the case $\delta=1$, where the ${\mathbb Z}_3$ rotational symmetry emerges, all the total energies are the same $E[K_b]=E[K_r]=E[K_g]=\frac{3\sqrt{3}}{4}$, and the energy lump associated with every solution is in this case indistinguishable in the space $x$.

\begin{figure}[h]
\centerline{\includegraphics[height=3.2cm]{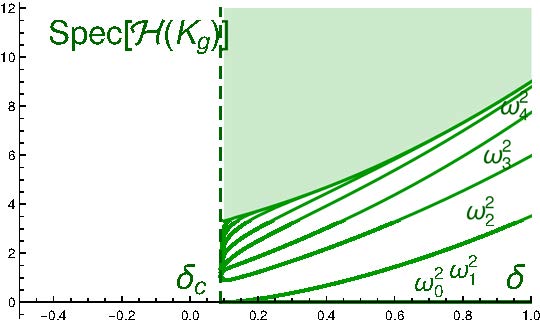} \hspace{0.3cm} \includegraphics[height=3.2cm]{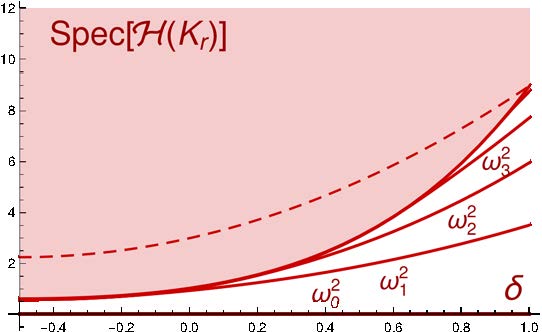} \hspace{0.3cm}  \includegraphics[height=3.2cm]{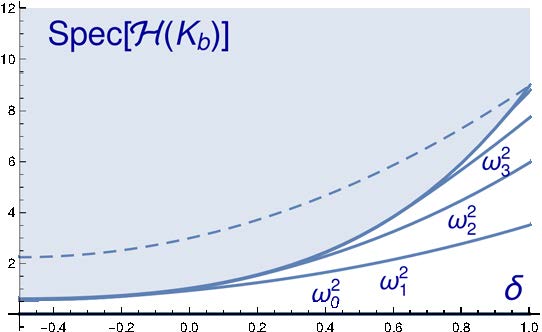}}
\caption{\small Graphics of the spectrum of the kink fluctuation operator (\ref{hessian}) for several values of $\delta$-parameter: (a) Green-kinks: ${\rm Spec} {\cal H}[K_g]$. (b) Red-kinks: ${\rm Spec} {\cal H}[K_r]$. (c) Blue-kinks: ${\rm Spec} {\cal H}[K_b]$.}\label{fig:spectrum1}
\end{figure}

The first order differential equation introduced in (\ref{complexedo}) has been employed to identify the static BPS kinks. However, the evolution of these solutions, for example in a scattering process, is ruled by the second order partial differential equations
\begin{eqnarray}
\frac{\partial^2 \phi}{\partial t^2} - \frac{\partial^2 \phi}{\partial x^2} &=& (\phi^3-(\delta-1)(\phi + \phi^2)-\delta)((\delta-1)(1+2\overline{\phi})-3\overline{\phi}^3)   \hspace{0.4cm},\label{pde1} \\
\frac{\partial^2 \overline{\phi}}{\partial t^2} - \frac{\partial^2 \overline{\phi}}{\partial x^2} &=& (\overline{\phi}^3-(\delta-1)(\overline{\phi} + \overline{\phi}^2)-\delta)((\delta-1)(1+2\phi)-3\phi^3)  \hspace{0.4cm}, \label{pde2}
\end{eqnarray}
obtained from the Euler-Lagrange equations associated with the action (\ref{complexaction}). With (\ref{pde1}) and (\ref{pde2}) the linear stability of the static BPS kinks $K_Q^{(i,j)}(x)$ can be studied by analyzing the evolution of a fluctuation of this solution. Plugging the form,
\[
\Psi(x,t)=\phi(x) + \epsilon\, e^{i\omega t}\, \psi(x) \ \ , \ \ \overline{\Psi}(x,t)=\overline{\phi}(x) + \epsilon\, e^{-i\omega t}\, \overline{\psi}(x)
\]
into (\ref{pde1}) and (\ref{pde2}), where $\phi(x) \equiv K_Q^{(i,j)}(x)$ is the scalar complex field of the BPS kink profile. Neglecting terms of order greater than 1 in $\epsilon$, the fluctuations $\psi(x)=\psi_1(x) + i \psi_2(x)$ must comply with the spectral problem
\begin{eqnarray}
&& -\frac{d^2}{d x^2} \psi(x) + V_{00}(x)\, \psi(x) + V_{01}(x)\, \overline{\psi}(x)  = \omega^2 \, \psi(x) \hspace{0.4cm}, \label{spectralproblem1} \\
&& -\frac{d^2}{d x^2} \overline{\psi}(x) + V_{10}(x) \, \overline{\psi}(x) + V_{11}(x) \, \psi(x)  = \omega^2 \, \overline{\psi}(x)\hspace{0.4cm}.  \label{spectralproblem2}
\end{eqnarray}
where the potential well components $V_{ij}(x)$, with $i,j=0,1$, are given by:
\begin{eqnarray*}
V_{00}(x) = V_{11}(x) &=& (3 \phi^2(x)-(\delta-1)(2\phi(x) + 1))(3 \overline{\phi}^2(x) - (\delta-1)(2 \overline{\phi}(x) + 1))   \hspace{0.5cm} , \\
V_{01}(x) = \overline{V_{10}(x)} &=&  2(\delta-\phi(x))(1 + \phi(x) + \phi^2(x))(\delta - 1 - 3 \overline{\phi}(x))  \hspace{2.8cm} .
\end{eqnarray*}
The kink fluctuation operator ${\cal H}[K^{(i,j)}(x)]$ is thus given by the second order differential matrix operator
\begin{equation} \label{hessian}
{\cal H}[K_Q^{(i,j)}(x)] = \left( \begin{array}{cc}  -\frac{d^2}{dx^2} + V_{00}(x) & V_{01}(x) \\ \overline{V_{01}(x)} & - \frac{d^2}{dx^2} + V_{00}(x) \end{array} \right)
\end{equation}
where $\phi(x)$ and $\overline{\phi}(x)$ must be numerically constructed. A numerical analysis of the spectral problem of ${\cal H}$ for the static $K_{b}(x)$, $K_{r}(x)$ and $K_{g}(x)$ kinks, as a function of the real $\delta$ parameter, is illustrated in Figure \ref{fig:spectrum1}. $\delta$  has been considered in the range $\delta > \delta_b$ due to the above commented mirror symmetry of the problem with respect to the $\delta=\delta_b$ value. As expected, the discrete spectrum in the three cases involves one zero mode $\omega_0^2$. The zero mode represents the fact that an infinitesimal translation of the kink solution, in the spatial coordinate, gives place to a new kink solution. The global behavior describes a kink moving with a certain velocity $v_0$.

The kinks $K_b(x)$ and $K_r(x)$ connect the vacuum $p_0$ with the vacua $p_1$ and $p_2$ respectively. In this case the matrix potential well asymptotically reaches different constant values. For example,
\[
\lim_{x\rightarrow -\infty} V[K_{r}(x)] =3(1+\delta+\delta^2) \, \mathbb{I}_{2\times 2} \hspace{0.4cm} \mbox{and} \hspace{0.4cm} \lim_{x\rightarrow \infty} V[K_{r}(x)] = (1+\delta+\delta^2)^2 \, \mathbb{I}_{2\times 2}
\]
For these cases, a doubly degenerate continuous spectrum emerges on the threshold value $\omega_c^2= (1+\delta+\delta^2)^2$. Another doubly degenerate continuous spectrum is immersed in the previous one on $\widetilde{\omega}_c^2=3(1+\delta+\delta^2)$. This level is graphically represented in Figure \ref{fig:spectrum1} (b) and (c) by a dashed curve. Besides, the number of discrete (shape) modes grows as the model parameter $\delta$ increases. On the other hand, the green kink $K_g(x)$ asymptotically goes from $p_1$ to $p_2$ giving rise to a symmetric potential well. Now a (fourfold) degenerate continuous spectrum emerges on the value $\omega_c^2= 3(1+\delta+\delta^2)$. In general, the discrete spectrum consists of several shape modes. However, a peculiar behavior occurs when the parameter $\delta$ tends to $\delta=\delta_c^+$. In this case the $K_g(x)$ approaches to a configuration of an antikink $K_{\overline{b}}(x)$ followed by an antikink $K_{\overline{r}}(x)$, which are infinitely separated, see Figure \ref{fig:Kinks3} (a). The potential well in this case increases its width as the value of $\delta$ tends to $\delta_c$. This process progressively captures new discrete modes, which are clustered at the parameter value $\delta_c$.

Notice that a static BPS kink can be transformed into a constant velocity traveling kink by using a Lorenz boost, that is,
\[
K_Q^{(i,j)}(x,t;v_0) = K_{Q \, {\rm static}}^{(i,j)}\Big(\frac{x-x_0-v_0 t}{\sqrt{1-v_0^2}} \Big)
\]
where $x_0$ is the kink center at $t=0$ and $v_0$ is a constant velocity. The kink center $x_C$ is moving at the same speed, $x_C=x_0+v_0 t$. The total energy of these traveling kinks is
\begin{equation}
E[K_Q^{(i,j)}(x,t;v_0)] = \frac{E[K_{Q\, {\rm static}}^{(i,j)}(x)]}{\sqrt{1-v_0^2}} \hspace{0.4cm} . \label{travelingenergy}
\end{equation}

\section{Kink-kink scattering}\label{sec3}

\begin{figure}[h]
	\centerline{\includegraphics[height=4.5cm]{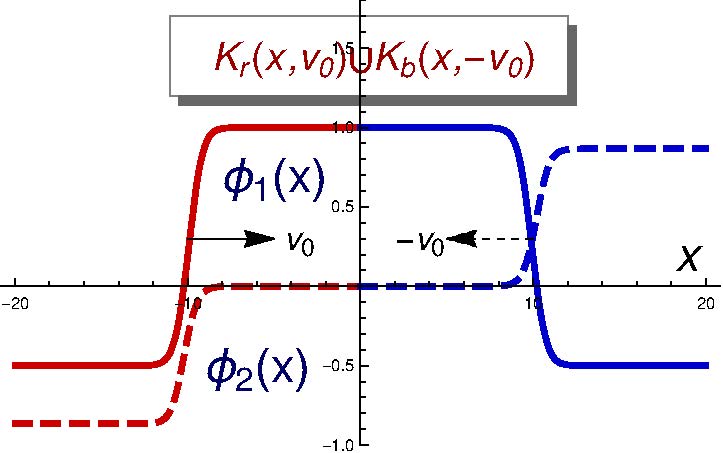} \hspace{0.6cm} \includegraphics[height=4.5cm]{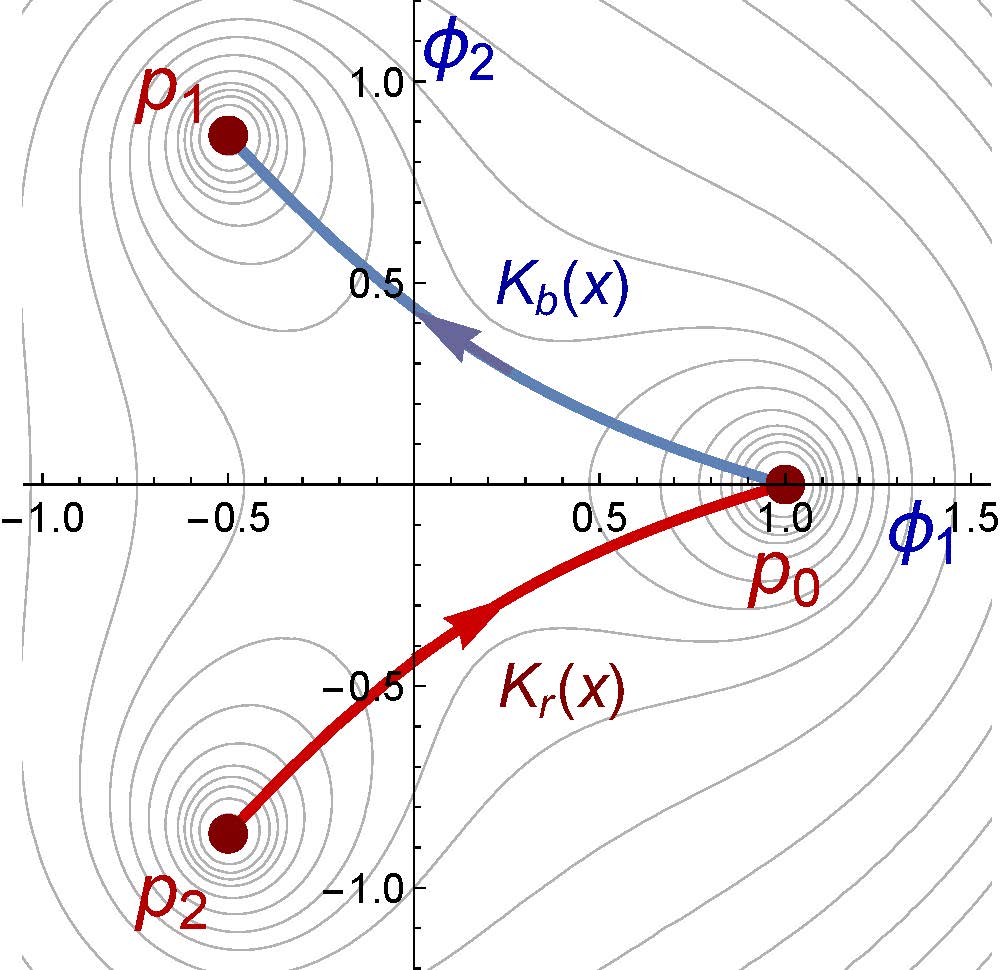} }
	\caption{\small (a) Profiles of the first and second field components associated to the initial configurations (\ref{concatenation1}) for the  kink-kink scattering processes, for the $\delta=1$ case. (b) Orbits of the initial configurations (\ref{concatenation1}) associated to the kink-kink scattering processes.} \label{fig:InitialConfiguration}
\end{figure}

The scattering processes between the two identical extended particles described in Section \ref{sec2} are discussed now. Recall that there exist essentially two regimes, $\delta \in (\delta_a,\delta_c)$ and $\delta \in (-\infty, \delta_a ) \cup (\delta_c,\infty)$, where either two or three kinks (and their corresponding antikinks) emerge. We shall focus in this work on analyzing the scattering between the $K_r$ and $K_b$ kinks as the value of the $\delta$-parameter varies. The initial configuration for our scattering problem is characterized by the concatenation
	\begin{equation}
	K_r (x+x_0,0;v_0) \cup K_b(x-x_0,0;-v_0) \hspace{0.5cm}, \label{concatenation1}
	\end{equation}
	where $x_0$ is large enough to guarantee the smoothness of the configuration (\ref{concatenation1}) and $i=0,1,2$. This configuration describes two well separated topological kinks, the first one asymptotically beginning at the vacuum $p_2$ and closely arriving to the vacuum $p_0$ at $x=0$; and the second one asymptotically going from this point to the vacuum $p_1$, see Figure \ref{fig:InitialConfiguration}. It can be checked that the first component of the multi-kink profile (\ref{concatenation1}) is an even function while the second one is an odd function. This behavior is preserved by the second order differential equations (\ref{pde1}) and (\ref{pde2}), so the kink components will evolve maintaining these symmetries. Notice that the antikink-antikink collisions $K_{\overline{r}} (x+x_0,0;v_0) \cup K_{\overline{b}}(x-x_0,0;-v_0)$ can be analyzed by means of the kink-kink scattering by simply employing the space mirror symmetry.

In the following subsections, we shall address the numerical study of the previously mentioned scattering processes. In order to accomplish this task a particular fourth-order explicit finite difference algorithm implemented with fourth-order Mur boundary conditions has been designed and applied to the non-linear Klein-Gordon equations (\ref{pde1}) and (\ref{pde2}) with initial configuration (\ref{concatenation1}). The details of this numerical method are explained in the Appendix. Note that the linear plane waves are absorbed at the boundaries in this numerical scheme avoiding that the radiation is reflected in the simulation contours.

A challenging issue in the kink scattering analysis in this two-component scalar field theory model corresponds to the identification of the evolving kinks at every time $t$. In some cases it is difficult to distinguish, for example, a vibrationally excited $K_g(x)$ green kink from the $K_r(x)$ and $K_b(x)$ kinks moving away. The orbits in these cases asymptotically begin and end, respectively, at the points $p_2$ and $p_1$ and their midpoints may be more or less near to the vacuum $p_0$. A convention on the value of this distance is used to distinguish both situations. On the other hand, a suitable way to determine the center of the traveling kinks must also be established. In the $\phi^4$ model, for instance, the location of the kink center is defined by the zero of the kink profile. In our model, however, the kink profile has two components, and it does not seem reasonable to bias the use of one of the components with respect to the other. In this framework, the kink center will be defined as the value of $x$ that determines the closest point of the evolving kink orbit to the midpoint of the corresponding static BPS orbit. Once the identities and the centers of the scattered kinks have been obtained, this information is used to work out its final velocities by employing a linear regression when the kinks are far enough apart from each other.

The study of the kink-kink scattering as the parameter value $\delta$ varies is carried out following several stages. First, the different scattering channels, which depend on the initial collision velocity, are identified. We shall consider the case $\delta=1$ to illustrate these processes. Secondly, the behavior of the final velocity $v_f$ of the scattered kinks as a function of the initial velocity $v_0$ is examined. This analysis allows us to obtain a global understanding of the the kink scattering in this model. Finally, the possible presence of resonance phenomena will be analyzed.

\subsection{Classification of the $K_r(x)$-$K_b(x)$ scattering channels} \label{sec31}

As previously mentioned the different scattering channels that arise in the $K_r(x)$-$K_b(x)$ collisions is described in this section. It has been found that there exist only two different outputs, which are described below:

\begin{enumerate}
		\item \textit{Kink-kink hybridization:} In this case, two well separated topological kinks $K_r(x)$ and $K_b(r)$ are pushed together with initial velocity $v_0$. This multikink configuration asymptotically starts at the vacuum $p_2$, crosses trough the vacuum $p_0$ and asymptotically arrives to the point $p_1$. The traveling kinks approach each other and finally collide. The midpoint of the evolving configuration orbit detaches from the vacuum $p_0$. The elastic forces approach the initial configuration to a more energetically favorable configuration going from $p_2$ directly to $p_1$. Thus, the collision between the red and the blue kink causes the annihilation of these kinks and the emergence of only one green kink pinned at $x=0$, as a consequence of the total momentum conservation. Therefore, the final velocity of this energy lump is zero, $v_f=0$. We have referred to these type of events kink-kink hybridization because the collision between two kinks gives place to a new different kink. Notice that the resulting kink belongs to a distinct topological sector from those comprising the colliding kinks. This class of scattering events can be symbolized as
		\begin{equation}
				K_r(v_0) \cup K_b(-v_0) \rightarrow K_g^{*}(0)+ \nu \label{kktokprocess}
		\end{equation}
		where the asterisk superscript in (\ref{kktokprocess}) stands for the excitation of the internal vibrational modes of the kink and $\nu$ represents radiation emission. This type of processes demands the existence of the $K_g(x)$-kink, so they can arise only for $\delta>\delta_c$ and $\delta<\delta_a$.
		
		A particular process (\ref{concatenation1}) for $\delta=1$ and initial velocity $v_0=0.2$ is illustrated in Figure \ref{fig:collision04kktok}, where the evolution of the field components is displayed. Besides, both the vibrational excitation and the radiation emission can be clearly observed in these two graphics.
	
	\begin{figure}[h]
		\centerline{\includegraphics[height=3.5cm]{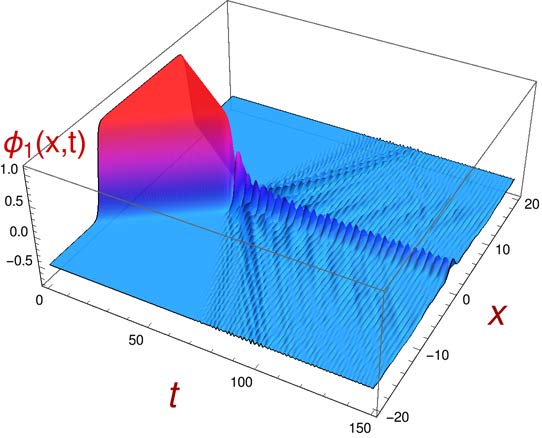} \hspace{1.5cm} \includegraphics[height=3.5cm]{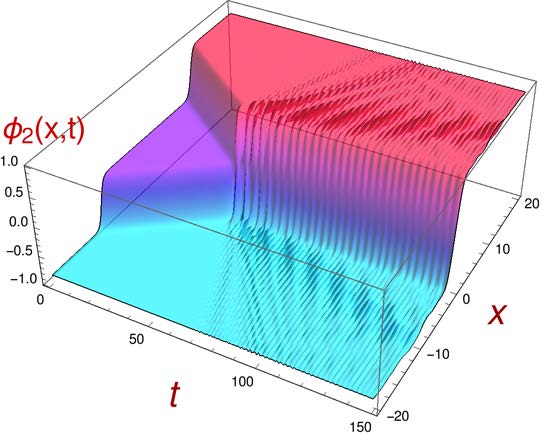}  }
		\caption{\small Graphical representation of a kink-kink scattering process with $\delta=1$ and impact velocity $v_0=0.2$. Evolution of the first and second field components for the $K_r(v_0) \cup K_b(-v_0) \rightarrow K_g^{*}(0)+ \nu$ scattering process. } \label{fig:collision04kktok}
	\end{figure}

	In Figure \ref{fig:collision04kktokorbits} the orbits traced by the solution of the equations (\ref{pde1}) and (\ref{pde2}) associated with the previous scattering event are plotted for different times. It can be observed how the initial configuration formed by the concatenation of two single BPS kink orbits living in two different topological sectors changes into one vibrating single BPS kink orbit living in the third topological sector.
	
	\begin{figure}[h]
		\centerline{\includegraphics[height=3cm]{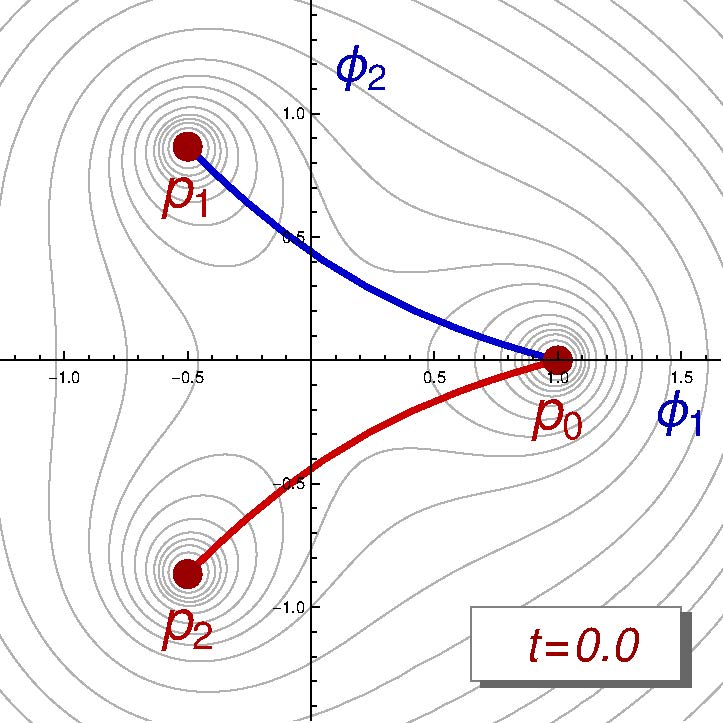} \hspace{0.5cm} \includegraphics[height=3cm]{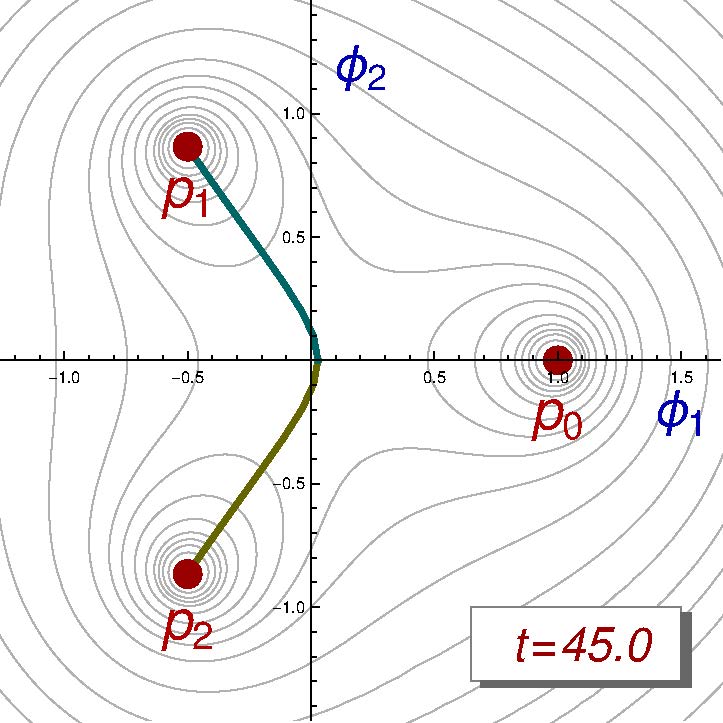} \hspace{0.5cm} \includegraphics[height=3cm]{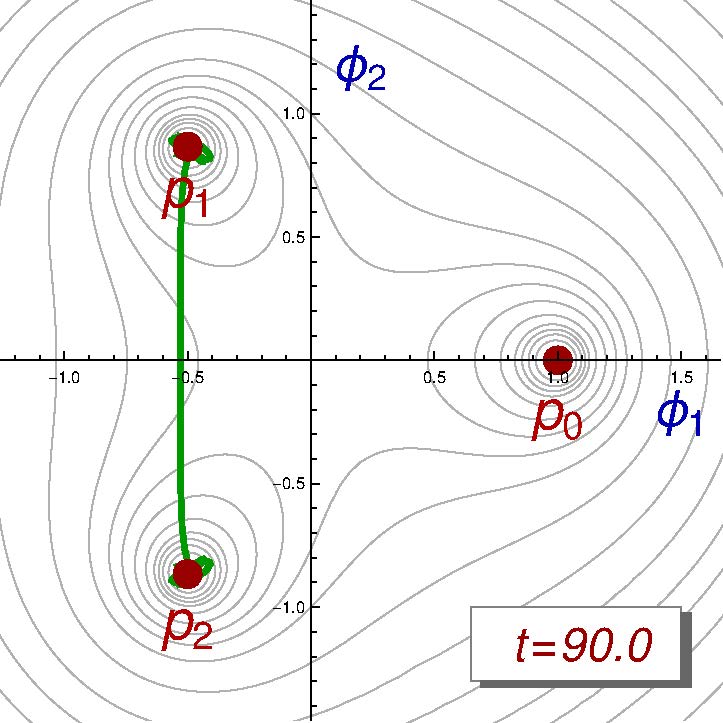}  \hspace{0.5cm} \includegraphics[height=3cm]{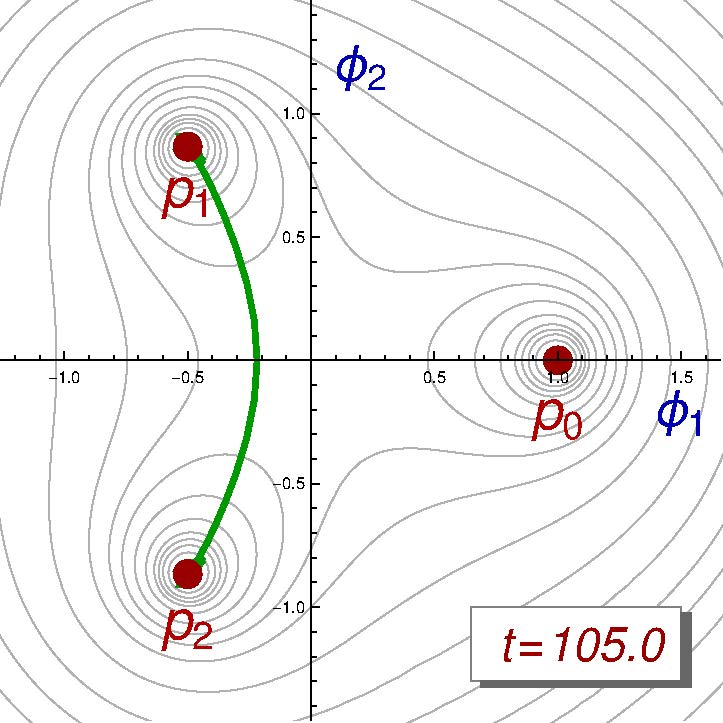} }
		\caption{\small Graphical representation of a kink-kink scattering process with $\delta=1$ and impact velocity $v_0=0.2$. Evolution of the orbit for the $K_r(v_0) \cup K_b(-v_0) \rightarrow K_g^{*}(0)+ \nu$ scattering process for several times.} \label{fig:collision04kktokorbits}
		\end{figure}
	
	From an energetic point of view, the initial configuration carries a total energy $E_0= 2 E[K_{r,b}(x,t;v_0)]$, see (\ref{travelingenergy}), whereas the static BPS kink energy is given by (\ref{energybr}) and (\ref{energyg}). This implies that the energy accumulated in the vibrational modes and emitted as radiation in this type of scattering events amounts to $2 E[K_{r,b}(x,t;v_0)]-E[K_{g}(x)]$. Therefore, at least the energy of a static BPS kink is transferred to these modes. Besides, this relocated energy is an increasing function of the collision velocity $v_0$, as expected.

	\item \textit{Kink-kink reflection:} Here, the two traveling kinks approach each other with initial velocity $v_0$, collide and bounce back with a certain final velocity $v_f$. Obviously, the final velocities of the scattered kinks are equal although with opposite directions. These processes can be characterized as
	\begin{equation}
	K_r(v_0) \cup K_b(-v_0) \rightarrow K_r^*(-v_f) \cup K_b^*(v_f) + \nu \hspace{0.5cm} . \label{kktokkprocess}
	\end{equation}
	The final velocity $v_f$ will be less than the initial velocity $v_0$ because part of the energy is employed to excite the vibrational modes of the scattered kinks and to emit radiation.
	
	\begin{figure}[h]
		\centerline{\includegraphics[height=3.5cm]{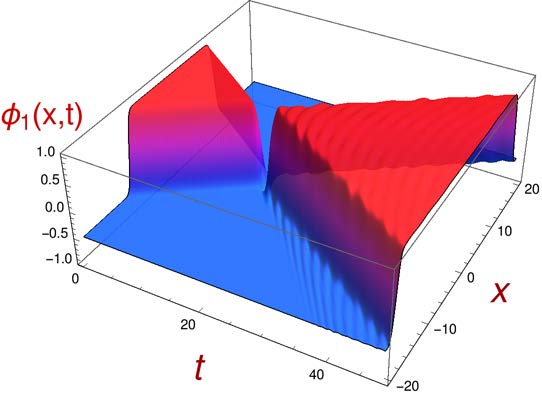} \hspace{1.5cm} \includegraphics[height=3.5cm]{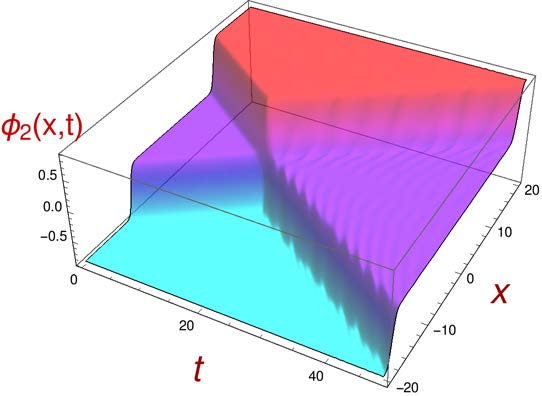}   }
		\caption{\small Graphical representation of a kink-kink scattering process for $\delta=1$ with impact velocity $v_0=0.6$. Evolution of the first and second field components for the $K_r(v_0) \cup K_b(-v_0) \rightarrow K_r^{*}(-v_f) \cup K_b^{*}(v_f)+ \nu$ scattering process. } \label{fig:collision06kktokk}
	\end{figure}

	The specific event $K_r(v_0) \cup K_b(-v_0) \rightarrow K_r^{*}(-v_f) \cup K_b^{*}(v_f)+ \nu$ with $v_0=0.6$ is displayed in Figure \ref{fig:collision06kktokk} for the parameter value $\delta=1$. Here, a red and a blue kink approach each other, collide, bounce back and, finally, move away with final velocity $v_f\approx 0.5669$, while vibrating and emitting radiation. The evolution of the multi-kink configuration orbit associated with this process is illustrated in Figure \ref{fig:collision06kktokkorbits}. The orbit of the initial configuration formed by the concatenation between the BPS kinks $K_r(v_0)$ and $K_b(-v_0)$ evolves approaching to a $K_g(v_0)$-kink orbit; however, the kinetic energy of the evolving kinks is large enough to make them to come back to the original kink-kink configuration. The kink orbits fluctuate because of the excitation of the vibrational modes. However, the energy transfer to these modes and the radiation emission is much less than in the previous type of scattering events. In this case, it becomes $2 E[	K_{r,b}(v_0)]-2 E[	K_{r,b}(v_f)]$, so only the difference between the initial and final kinetic energies is employed to excite the resulting kinks.

	\begin{figure}[h]
		\centerline{\includegraphics[height=3cm]{KSWZ_Fig10a} \hspace{0.5cm} \includegraphics[height=3cm]{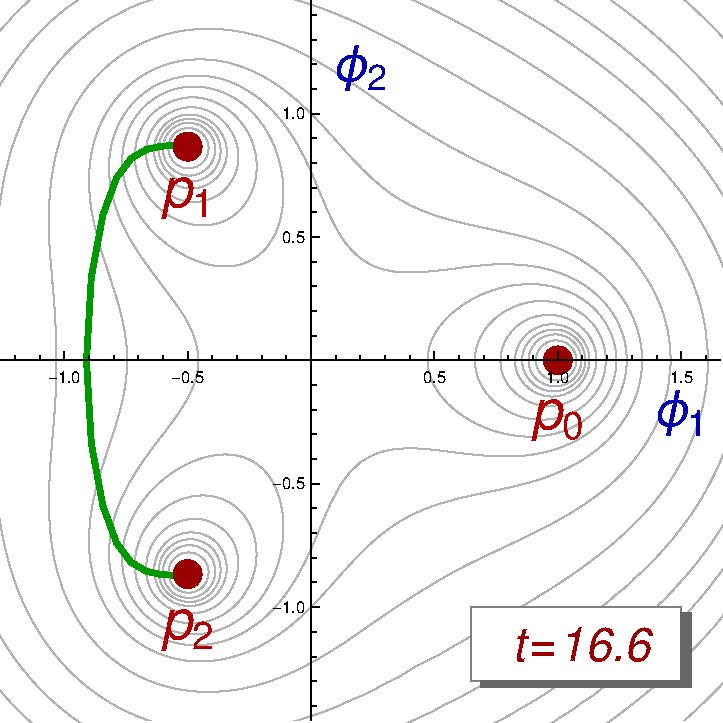} \hspace{0.5cm} \includegraphics[height=3cm]{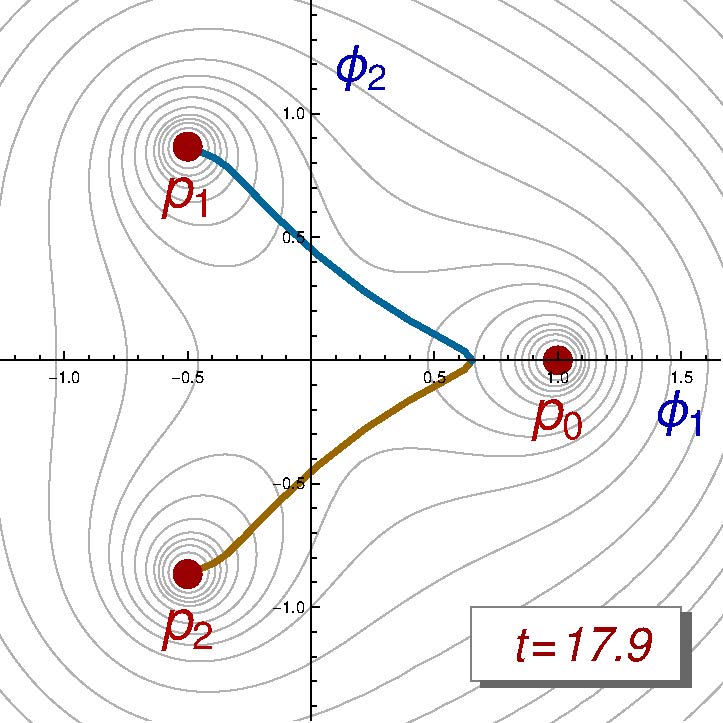}  \hspace{0.5cm} \includegraphics[height=3cm]{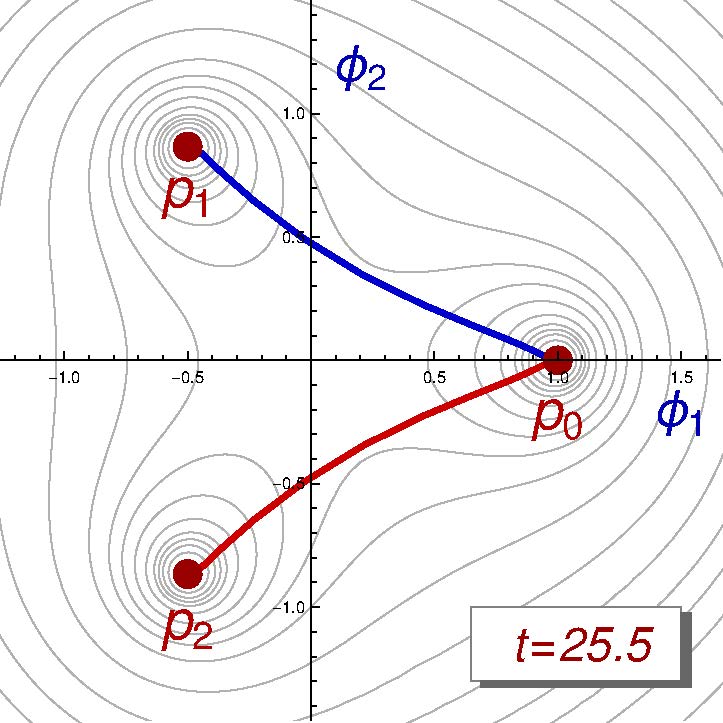} }
		\caption{\small Graphical representation of a kink-kink scattering process for $\delta=1$ with impact velocity $v_0=0.6$. Evolution of the orbit for the $K_r(v_0) \cup K_b(-v_0) \rightarrow K_r^{*}(-v_f) \cup K_b^{*}(v_f)+\nu$ scattering process for several times.} \label{fig:collision06kktokkorbits}
	\end{figure}
	
\end{enumerate}

In sum, there exist two distinct scattering channels when two symmetric kinks (or antikinks) collide. In the first case, the collision between two identical extended particles with different flavors gives place to only one new particle endowed with a new flavor, distinct to those of the colliding objects (\textit{kink-kink hybridization}). In the second case, the two energy lumps collide and reflect each other (\textit{kink-kink reflection}). The scattering between two kinks with different colors never leads to the annihilation of the topological solutions because these configurations start and end at different vacua. Notice, however, that the collision of three kinks with different colors can cause the annihilation of the three extended particles.

\subsubsection{Final versus initial velocity diagrams}

\begin{figure}[h]
\centering\begin{tabular}{c}
\includegraphics[height=3cm]{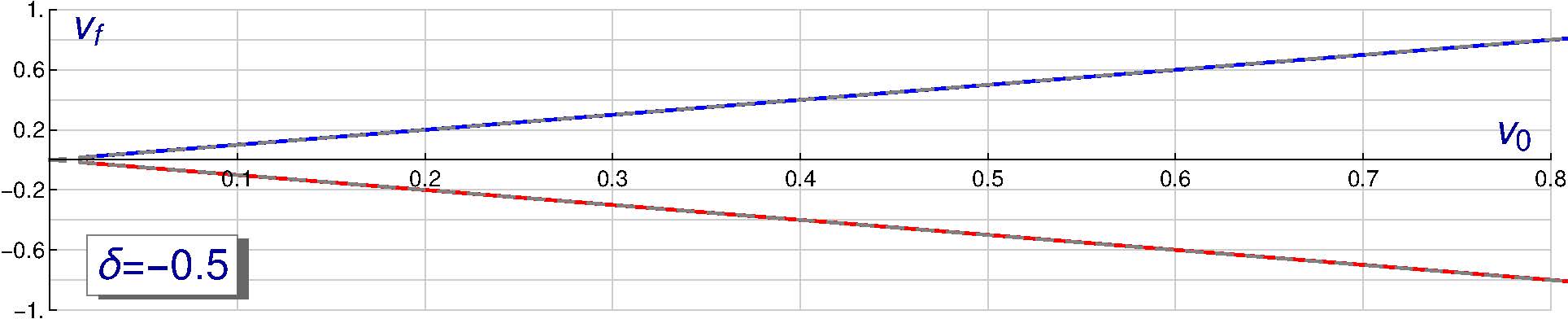} \\
\includegraphics[height=3cm]{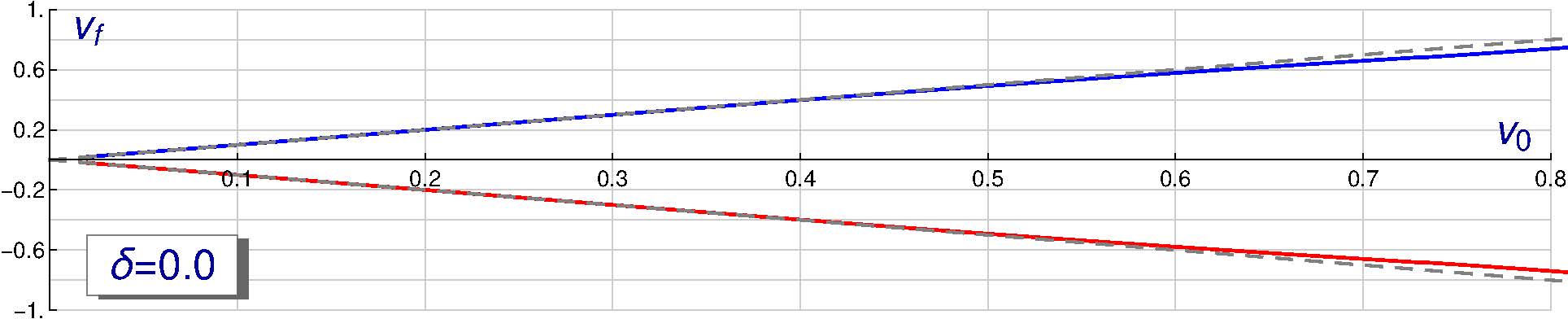} \\
\includegraphics[height=3cm]{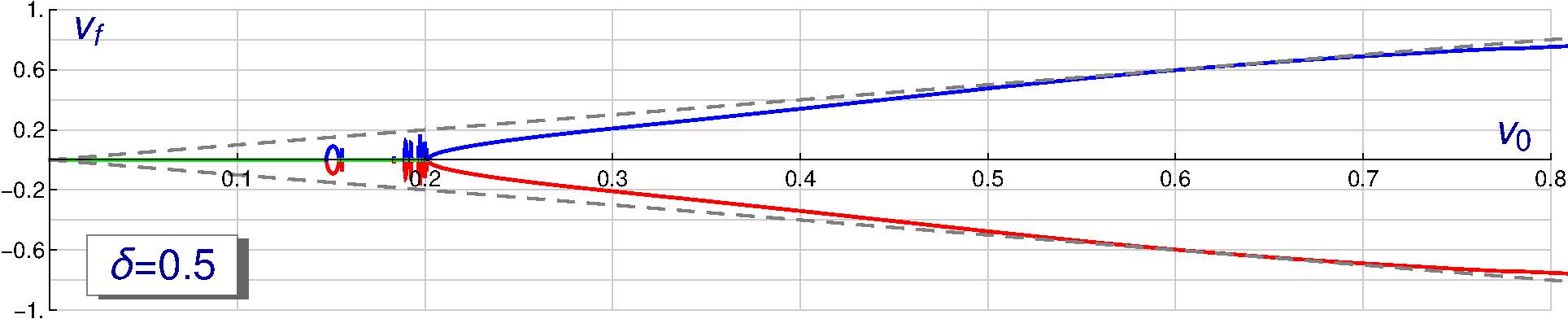} \\
\includegraphics[height=3cm]{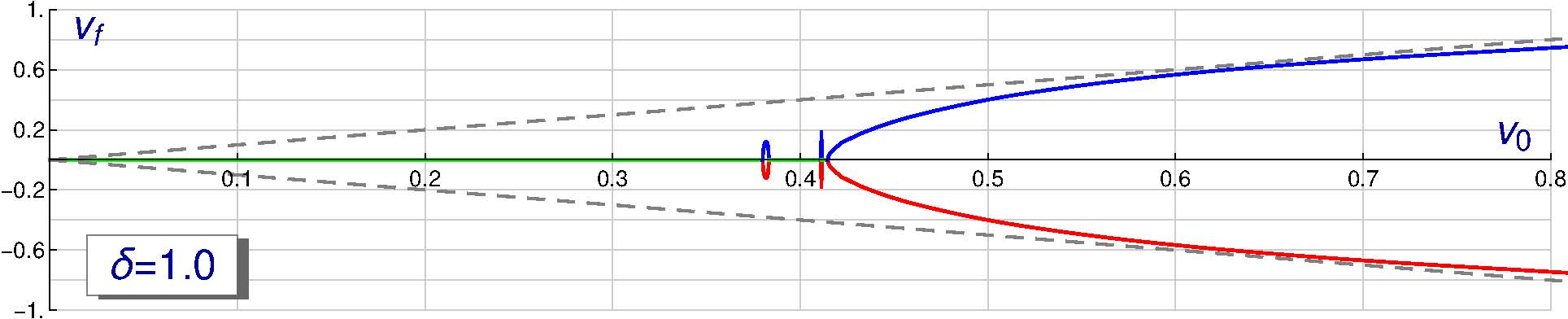}\end{tabular}
\caption{\small Graphical representation of the final velocity $v_f$ of the kinks as a function of the initial velocity $v_0$ for the $K_r(x)$-$K_b(x)$ scattering processes for the cases: (a) $\delta=-0.5$, (b) $\delta=0.0$, (c) $\delta=0.5$ and (d) $\delta=1.0$. Green, red and blue curves  determine, respectively, the final velocities of the $K_g(x)$-kinks, the $K_r(x)$-kinks and the $K_b(x)$-kinks.} \label{fig:diagramavelocidadkinkkink}
\end{figure}

In this section a global insight of the scattering problem in this type of events is aimed. For this purpose we shall study the dependence of the previously introduced scattering channels on the initial velocities $v_0$ of the colliding kinks and on the model parameter $\delta$. Numerical simulations have been massively carried out changing the initial collision velocity $v_0$ with a velocity step $\Delta v_0=0.001$, which is decreased to $\Delta v_0=0.0001$ in the resonance regime, which will be discussed later. The model parameter is varied with step $\Delta \delta = 0.1$. The compilation of the data can be visualized in Figure \ref{fig:diagramavelocidadkinkkink}, where the final velocity $v_f$ of the scattered kinks as a function of the initial velocity $v_0$ is graphically represented for several representative values of $\delta$. In Figure \ref{fig:diagramavelocidadkinkkink} the colors of the curves are used to identify the scattered kinks: a green line is employed to determine the final velocity of the green kinks, a red curve defines the final $K_r(x)$-kink velocity whereas a blue curve is used for the blue kinks. Following this convention, it is clear, for example, that for the case $\delta=1.0$ the collision between two kinks with initial velocity $v_0=0.2$ gives place to a hybrid kink whereas if $v_0=0.6$ the two kinks reflect each other.

The velocity diagram for $\delta=-1/2$ shows that the scattering processes are practically elastic for all the values of the initial velocity. In this case the dynamics of the model reduces to a one-dimensional $\phi_2^6$ model in the ordinate axis of the internal plane. The three vacua are aligned in the straight line $\phi_1=-\frac{1}{2}$, such that the first components of the red and blue kinks are constant and remain decoupled from the dynamics. This means that the orthogonal $\phi_1$-fluctuations do not affect the kink evolution. Only the kink reflection channel takes place for this particular value of $\delta$. Notice that the green kink is not present in this case because $\delta=\delta_b < \delta_c$. A similar behavior happens for $\delta=0.0$ although here for large impact velocities the events are clearly not elastic. In this case the vacuum $p_0$ leaves the aligned location and the effects of fluctuations in the $\phi_1$-direction are apparent. The pattern dramatically changes for $\delta > \delta_c$. In this case the green kinks emerge in the model and kink-kink hybridization is now possible.

The pattern found in Figure \ref{fig:diagramavelocidadkinkkink} for the cases $\delta=\frac{1}{2}$ and $\delta=1$ reveals that the kink-kink hybridization regime is predominant for small initial velocities $v_0$. An unbroken kink reflection window arises from approximately the threshold value $v_0=0.1474$ for $\delta=\frac{1}{2}$ and $v_0=0.4146$ for $\delta=1$. Indeed, this regime is more prevalent as the $\delta$ increases. Clearly, kink reflection is more energetically demanding than kink-kink hybridization because it involves two scattered kinks instead of one. As a consequence, reflection occurs when the kinetic energy carried by the kinks is large enough.

\begin{figure}[h]
	\centerline{\includegraphics[height=1.5cm]{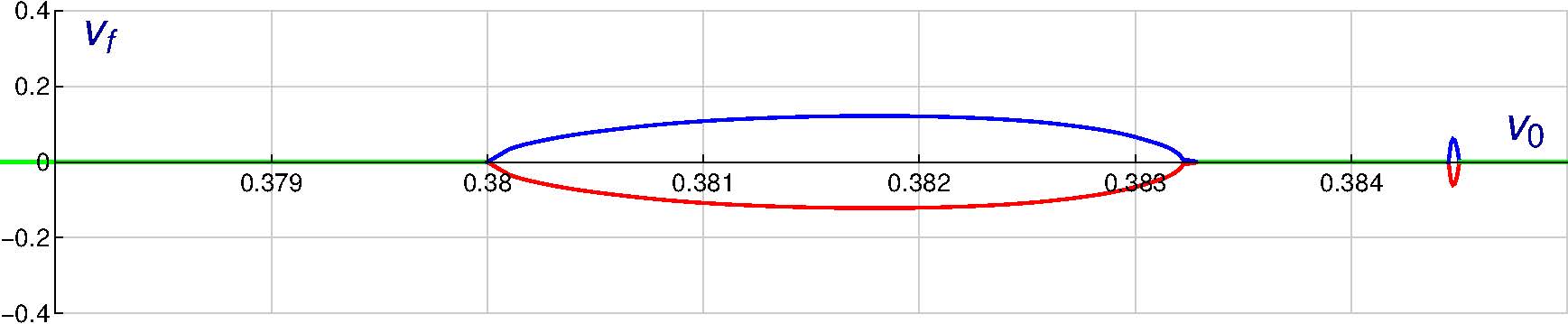} \hspace{0.5cm} \includegraphics[height=1.5cm]{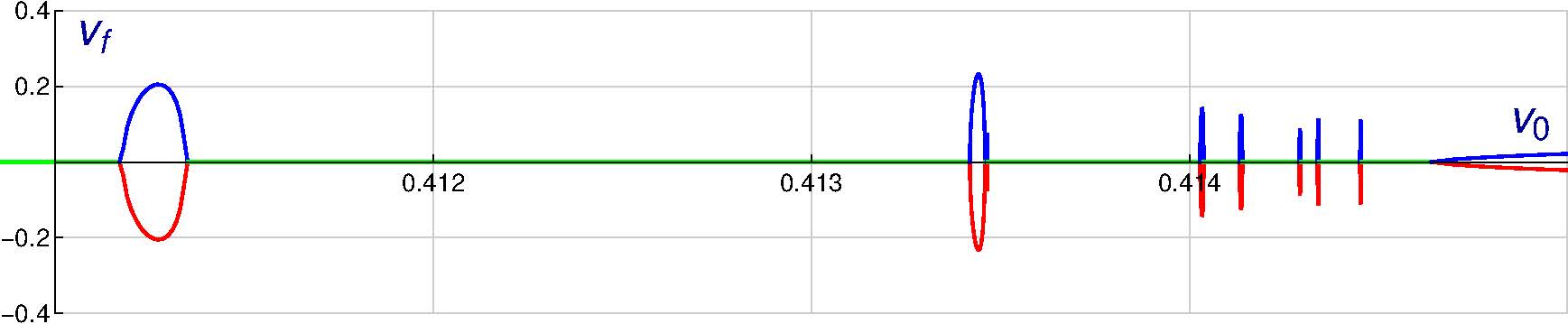}}
	\caption{\small Resonance windows for the $K_r(x)-K_b(x)$ scattering processes for the case $\delta=1$. Green, red and blue curves  determine, respectively, the final velocities of the $K_g(x)$-kinks, the $K_r(x)$-kinks and the $K_b(x)$-kinks.} \label{fig:diagramavelocidadkinkkinkb}
\end{figure}

In the transition between these two previously described scenarios there exist isolated windows where the two previous regimes are interlaced, see Figure \ref{fig:diagramavelocidadkinkkink} (c) and (d). These windows have been enlarged in Figure \ref{fig:diagramavelocidadkinkkinkb} for the case $\delta=1$. The first and widest of these bands in the case $\delta=1$ is approximately defined in the interval $[0.380,0.3832]$. Successively narrower windows appear as we approach to the threshold value $v_0=0.4146$. As usual, the phenomenon behind this behavior is the \textit{resonant energy transfer mechanism}. Energy can be exchanged between the zero and the vibrational modes of the kinks in the collision.

\begin{figure}[h]
	\centerline{\includegraphics[height=3.5cm]{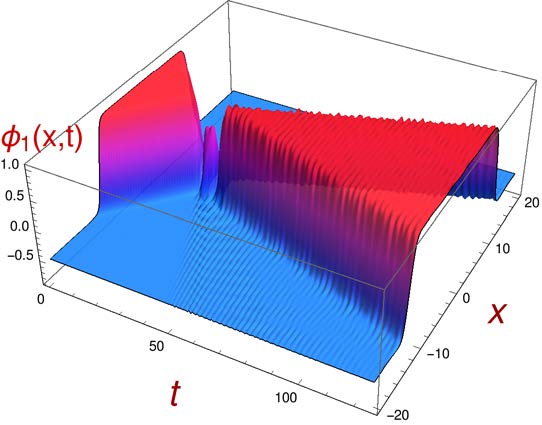} \hspace{1.5cm} \includegraphics[height=3.5cm]{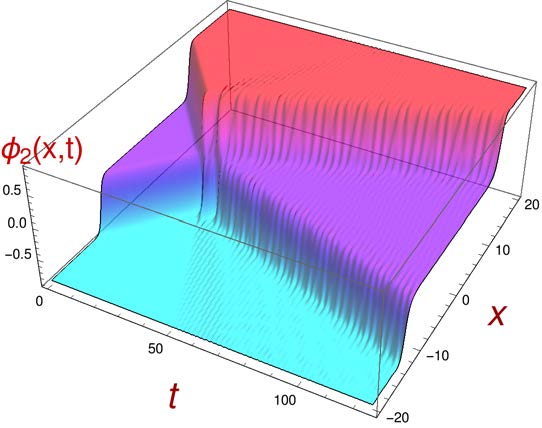}  }
	\caption{\small Graphical representation of a kink-kink scattering process with $\delta=1$ and impact velocity $v_0=0.382$. Evolution of the first and second field components for a $K_r(v_0) \cup K_b(-v_0) \rightarrow K_r^{*}(-v_f) \cup K_b^{*}(v_f)+\nu$ kink scattering event with two bounces.} \label{fig:collision0382kktokk}
\end{figure}

This scenario has been illustrated in Figure \ref{fig:collision0382kktokk}, where the scattering between two kinks with initial velocity $v_0=0.382$ is plotted for $\delta=1$. This value of the collision velocity belongs to the first resonance windows, as previously mentioned. Here, a red and a blue kinks approach each other with speed $v_0=0.382$ and collide. For a period of time a green kink arises whose vibrational modes are highly excited. This vibration causes the emergence of the original kinks, but they cannot escape because of the loss of kinetic energy. The attraction force makes them to approach again and collide a second time. The same procedure takes place but, on this occasion, energy from the vibrational modes is transferred to the zero mode, increasing the kinetic energy of each kink. In this particular event, this rise is enough to let the kinks escape and move away with a traveling velocity $v_f=0.120875$, see Figure \ref{fig:collision0382kktokk}. Therefore, in this scattering process kinks suffer two bounces.

As observed in Figure \ref{fig:diagramavelocidadkinkkinkb}, the number of resonance intervals is not too large and, in general, its width is very small. It is assumed the existence of other windows although their length is too small to be detected. From our point of view the scarce presence of the resonance windows in this model compared with the $\phi^4$-model is related to the existence of several kink vibrational modes. The resonant energy transfer mechanism can operate in several modes, so the probability that the energy previously transferred from the zero mode to the vibrational modes return to that mode again is less than in the $\phi^4$ model, where there are only two involved modes.

\section{Conclusions and further comments} \label{sec4}

In this paper, the kink scattering in the dimensional reduction of the bosonic sector of a generalized Wess-Zumino model, where the vacua are located at the vertices of an isosceles triangle, is analyzed. The specific vacuum distribution is set by the value of the model parameter $\delta$ arising in the potential term (\ref{potential}). Two special cases in this context are given by the value $\delta=-\frac{1}{2}$, where the three vacua are aligned and an embedded $\phi^6$ model rules the dynamics, and the value $\delta=1$, where the model has a three-fold rotational symmetry and the vacua are placed at the vertices of an equilateral triangle. In general, two BPS kinks/antikinks (called red and blue) arise for any value of the model parameter $\delta$. These kinks are related by a $\phi_2$-reflection symmetry. A third different BPS kink/antikink (called green) emerges only for some ranges of the parameter. The kink scattering between the two symmetric kinks have been analyzed. When only the two symmetric kinks arise in the theory the only existing scattering channel is given by \textit{kink-kink reflection} (kinks collide, bounce back and move away). When the third kink is present a new possibility emerges: \textit{kink-kink hybridization} (kinks collide, annihilate and give rise to a new kink with a different color). In this regime, the last scattering channel is prevalent for low values of the initial velocity where as kink reflection prevails for large values. In the intermediate range some resonance windows emerge where $n$-bounce events take place.

In sum, the kink scattering in this two-component scalar field theory model comprise some novel processes, not previously described. This work opens new prospects for future research. From our point of view, it would be interesting to study collisions between kinks arising in other two-component scalar field theory models, and to analyze the relation between orthogonal and longitudinal eigenmodes with the resonance phenomenon.

\section{Acknowledgments}

The authors acknowledge the Junta de Castilla y Le\'on for financial help under grants BU229P18 and SA067G19, and Fundaci\'on Sol\'orzano Barruso under grant FS/25-2020. This
research has made use of the high performance computing resources of the Castilla y Le\'on Supercomputing Center (SCAYLE,
www.scayle.es), financed by the European Regional Development Fund. (ERDF).

\section{   Appendix   }\label{S:Appendix}

In this Appendix we shall describe the fourth-order finite difference numerical method with Mur conditions designed to study the kink scattering in the model introduced in this paper. For the sake of simplicity, this analysis is restricted to Klein-Gordon equations:
\begin{equation}
\label{eq_gral} \phi_{tt} = \phi_{xx}+F(\phi); \quad -\infty \leq x \leq \infty, \quad 0 \leq t
\leq t_e,
\end{equation}
where $F(\phi) = (F_1(\phi), F_2(\phi) )$ is a sufficiently differentiable function and $t_e \in \mathbb{R}^+$. Note that subscripts are used to denote partial derivatives. The initial conditions are determined by the expressions
\begin{eqnarray}
\label{init1} \phi(x,0)=f(x)=(f_1(x), f_2(x)), \quad -\infty \leq x \leq \infty, \\
\label{init2}  \phi_t(x,0)=\bar{f}(x)= ( \bar{f_1}(x), \bar{f_2}(x) ), \quad -\infty \leq x \leq \infty.
\end{eqnarray}
Since we would like to use numerical methods in our simulations, we consider this problem in a large interval $[-s, s]$ instead of $(-\infty, \infty)$, and we will develop appropriate (through high-order formulae) Mur conditions to approximate the solutions in the boundary $x=-s$, and $x=s$. We consider a uniform mesh of the region $\Omega = [-s,s]
\times [0,t_e]$. The vertices of the mesh will be $( x_i, t^n)$
where $t^n=n k$, $n=0,1, \ldots, N$, with $k=\Delta t$, and $x_i= -s + i h$, $i=0,1, \ldots, M=(2 s)/h$, with $h=\Delta x$.  We will denote $n_c= \frac{k}{h}$, $F_{i}^{n}$ the approximation of $ F(x_i, t^n) $ and $ \phi_{i}^{n} $ the numerical approximation of  $ \phi(x_i, t^n) $.
The numerical schemes built in this paper consist on three or more levels in time, therefore it is necessary to use another procedure to obtain an adequate approximation for the first step. We shall detail the algorithms used for every iteration, which involve explicit and implicit schemes and, finally, we shall address the boundary conditions.

\begin{enumerate}
	\item \textit{First iteration.} For the first iteration, we merely consider
\begin{equation} \label{caso1}
\begin{split}
&\phi_i^1 = \phi_i^0 + k (\phi_t)_i^0 + \frac{k^2 (\phi_{tt})_i^0}{2}+ \frac{k^3
	(\phi_{ttt})_i^0}{6} + \frac{k^4 (\phi_{tttt})_i^0}{24}  + O(k^5)
\end{split}
\end{equation}
to build a fourth-order approximation. $\phi_i^0$ and $(\phi_t)_i^0$ are
known from the initial conditions \Eqref{init1} and \Eqref{init2}.  Additionally, since $\phi_{tt} = \phi_{xx}+F(\phi)$, then $(\phi_{tt})_i^0 =
f''( -s + ih)  + F( \phi_i^0) $. In a similar way $(\phi_{ttt})_i^0$,
$(\phi_{tttt})_i^0$ and $(\phi_{ttttt})_i^0$ are obtained. Obviously,
\Eqref{caso1} becomes a different equation depending on the
nonlinear Klein--Gordon Equation (according to $F(\phi)$ and initial conditions).

\item \textit{Second to fourth iteration: implicit scheme.} After the first iteration, we are ready to use our finite difference methods.  Unfortunately, our explicit scheme (described below in equation (\ref{m_expl_4}))) has six levels.  Therefore, it is necessary to use an implicit algorithm before.  The central implicit method is as follows:
\begin{eqnarray}
\phi_i^n &=&  2 \phi_i^{n-1} -  \phi_i^{n-2}   +
n_c^2 ( \phi_{i-1}^{n-1}    - 2  \phi_{i}^{n-1} + \phi_{i+1}^{n-1}  )  + \nonumber \\ &  + & (n_c^2 - 1) n_c^2  \left( \frac{ 6  \phi_{i}^{n-1}   + \phi_{i-2}^{n-1}  - 4 \phi_{i-1}^{n-1}  -
	4 \phi_{i+1}^{n-1}  +  \phi_{ i+2 }^{ n-1 }  } { 12 } \right)  +  \label{m_impl_4}\\ &+   &
k^2 \left(  F_{i}^{n-1}  +  \frac{  F_{i}^{n}  - 2 F_{i}^{n-1}  + F_{i}^{n-2} }  {12}  \right.
+  \left. \frac{   n_c^2  ( F_{i-1}^{n-1}  - 2  F_{i}^{n-1}  +  F_{i+1}^{n-1} ) }  {12 }   \right) \nonumber
\end{eqnarray}
where $n=2, 3, 4$ and $i =2, \ldots, M-2$.

For the case $n=2, 3, 4$ and $i=1$, we have to replace in (\ref{m_impl_4}) the five-term central (in space) approximation
\[  \left( \frac{ 6  \phi_{i}^{n-1}   + \phi_{i-2}^{n-1}  - 4 \phi_{i-1}^{n-1}  -
	4 \phi_{i+1}^{n-1}  +  \phi_{ i+2 }^{ n-1 }  } { 12 h^2 } \right),  \]
by a forward (in space) formula:
\[       -   \left( \frac{ 9  \phi_{i}^{n-1}  +   \phi_{i+4}^{n-1}  -
	2 (3  \phi_{ i+3 }^{n-1} - 7  \phi_{i+2}^{n-1}  + 8  \phi_{i+1}^{n-1}  +
	\phi_{i-1}^{n-1}  )  }    { 12  h^2 }    \right).    \]
Similarly, in the case $n=2, 3, 4$ and $i=M-1$, we replace the five-term central approximation in (\ref{m_impl_4}) by the symmetric backward formula.

The finite difference method given by equation in (\ref{m_impl_4})  is expressly constructed to approximate (\ref{eq_gral}) with fourth-order, adding terms to remove the first terms of the local truncation error of the well-known method
\[  \frac{   \phi_i^n -
	2 \phi_i^{n-1} +  \phi_i^{n-2}   } { k^2 }  =   \frac{  \phi_{i-1}^{n-1}    - 2  \phi_{i}^{n-1} + \phi_{i+1}^{n-1}  }  { h^2 } +  F_{i}^{n-1},     \]
thus it is easy to check that its leading truncation error  is (whenever $k = n_c h$):
\[   \frac{  n_c^4 \left(2 \phi^{( 0, 6)} ( x_i,t^n) -
	5 F^{( 0, 4)} ( x_i, t^n) \right)   }  {  720  } h^4    -
\frac{
	5 n_c^2 F^{(4, 0)} ( x_i, t^n) + \left( 8 -
	10 n_c^2 \right) \phi^{(6, 0)} (x_i, t^n)  }  {  720  } h^4 + O(h^5).       \]

\item \textit{Following iterations: Explicit scheme.} In most of the iterations we prefer using an explicit, and much cheaper algorithm.  Since we want to obtain an accurate, but also stable scheme, after several tests, we finally decided to replace third lines in the implicit method, equation (\ref{m_impl_4}), by
\[  + k^2  \left(   F_{i}^{n-1} + \frac{   35 F_{i}^{n-1} + 11 F_{i}^{n-5} - 56 F_{i}^{n-4} +
	114 F_{i}^{n-3} - 104 F_{i}^{n-3} } {144}  +   n_c^2   \frac{   F_{i-1}^{n-1} - 2  F_{i}^{n-1}   +   F_{i+1}^{n-1} }  {12}   \right),   \]
thus, the explicit algorithm can be written as
\begin{eqnarray}
\phi_i^n &=&
2 \phi_i^{n-1} -  \phi_i^{n-2}   +
n_c^2( \phi_{i-1}^{n-1}    - 2  \phi_{i}^{n-1} + \phi_{i+1}^{n-1}  )   + \nonumber \\  &+& (n_c^2 - 1) n_c^2 \left( \frac{ 6  \phi_{i}^{n-1}   + \phi_{i-2}^{n-1}  - 4 \phi_{i-1}^{n-1}  -
	4 \phi_{i+1}^{n-1}  +  \phi_{ i+2 }^{ n-1 }  } { 12 } \right)  + \nonumber \\
&+&   k^2 \left(  F_{i}^{n-1} + \frac{   35 F_{i}^{n-1} + 11 F_{i}^{n-5} - 56 F_{i}^{n-4} +
	114 F_{i}^{n-3} - 104 F_{i}^{n-3} }  {144}  + \right. \label{m_expl_4}\\ && \left. \hspace{0.8cm}+\frac{   n_c^2  ( F_{i-1}^{n-1}  - 2  F_{i}^{n-1}  +  F_{i+1}^{n-1} ) }  {12 }   \right), \nonumber
\end{eqnarray}
with $n=5, \ldots, N$ and $i =2, \ldots, M-2$. For the cases $i=1$, and $M-1$, we replace the five-term central formula (in the second line in (\ref{m_expl_4})) by the forward and backward formulae given in the previous point.

As before, this explicit method was derived by adding terms to cancel the first terms of the truncation error, thus the leading truncation error of this method is
\[   \frac{ -5 n_c^2  F^{( 4, 0)} ( x_i,t^n) +2 \left(
	(4 - 5 n_c^2) \phi^{( 6, 0)} ( x_i, t^n)  +  n_c^4  \phi^{( 0, 6)} (x_i, t^n)  \right)  }  {  720  } h^4 + O(h^5).       \]

We tried to develop explicit finite-difference methods with one level less than the one given by  equation (\ref{m_expl_4})  in a similar way to those methods developed and analyzed in \cite{EG16,MV18}, but there $n_c$ was established as $1$, and for stability purposes now we utilized $n_c<1$.  Other numerical schemes proposed for similar nonlinear Klein-Gordon (or similar) hyperbolic partial differential equations are described in \cite{Lyn99,KhSa10,RaGhJa10} and references therein, they are mostly based on finite difference algorithms in time and finite difference or spectral methods in space. Unfortunately we were not able to derive an explicit fourth-order similar scheme with ``good'' stability properties. Numerically, we can check that both algorithms, the explicit and the implicit ones, are stable whenever $n_c=k/h < 1$ and they are employed to solve the wave equation, i.e., (\ref{eq_gral}) but replacing $F(\phi)$ by $F(x,t)$.   To do it, one can just check that both conditions in \cite[Thm 8.2.1]{Str04} are satisfied if $n_c=k/h < 1$.  It is necessary first to obtain the Von-Neumann amplification polynomial $\Phi(g,\theta)$ associated to both methods, calculating the maximum in absolute value of the roots of  this polynomial and studying when this is smaller than $1$. Since Klein-Gordon problems are nonlinear, it is not easy to study the stability (and therefore convergence) of the proposed methods.
To the best of our knowledge, the von Neumann stability analysis has not been
rigorously justified for nonlinear equations, but it is often justified approximately,
assuming that the solution $u(x, t)$ (and its numerical counterpart) does not vary
too rapidly for small $k, h$ values.  Although, numerically it looks that both methods are stable whenever $n_c=k/h < 1$ in the linearised version of our PDE;  however, since our problem is nonlinear, we suggest to use (and we employed) lower values for $n_c$ (for example $n_c \leq 0.5$), and moderate  $k, h$ values to avoid stability problems.

\item \textit{Mur conditions.} Equation (\ref{eq_gral}) is considered for $ x \in (-\infty, \infty)$. Since we want to solve it numerically, it is necessary to truncate the domain to $[-s, s]$ for $s$ large enough, and adding adequate boundary conditions.  In similar cases, Mur conditions (absorbing boundary conditions, ABC)  have traditionally been employed. Assuming that at $x_0=-s$, we can consider that $\phi_t (x_0,t)= \phi_x (x_0,t)$ (one-way wave equation, this option absorbs the wave and reflects as
little energy as possible).  And similarly at $x_M=s$, we consider that $\phi_t (x_M,t)= - \phi_x (x_M,t)$.
In some scientific papers, readers can found that at $(x_0, t^{n+1})$, we can calculate the numerical approximation as:
\begin{equation}  \label{Mur}
\begin{split}
\phi_0^{n+1} = \phi_1^{n+1} + \frac{ n_c-1} {n_c+1} (  \phi_1^{n+1} - \phi_0^{n}), \hspace{0.5cm} n=0, \ldots N-1,
\end{split}
\end{equation}
and the anti-symmetric condition at $(x_M, t^{n+1})$
\begin{equation}  \label{Mur2}
\begin{split}
\phi_M^{n+1} = \phi_{M-1}^{n+1} - \frac{ n_c-1} {n_c+1} (  \phi_{M-1}^{n+1} - \phi_M^{n}), \hspace{0.5cm}  \, n=0, \ldots N-1.
\end{split}
\end{equation}

However, these Mur conditions are only second-order.  Since we had developed fourth-order approximations for the central points, we also calculated new fourth-order boundary conditions:
\begin{equation}  \label{Mur4th}
\begin{split}
\phi_0^{n+1} =  - \frac{ 2 - n_c} {n_c+1} \phi_{1}^{n+1} + \frac{ (1-n_c) (2-n_c) } { 2(1+n_c) }  \phi_{0}^{n} + (2-n_c) \phi_1^{n} + \frac{n_c}{2} \phi_2^{n},\hspace{0.5cm} n=0, \ldots N-1,
\end{split}
\end{equation}
and similarly in the other border.

\end{enumerate}

\end{document}